\pdfoutput=1
\makeatletter
\def\input@path{{source/}}
\makeatother
\documentclass[aps,pra,twocolumn,10pt,superscriptaddress,longbibliography,floatfix,nofootinbib]{revtex4-2}
\usepackage{amsmath,amssymb,mathtools,amsthm}
\usepackage{graphicx}
\usepackage{booktabs}
\usepackage{enumitem}
\usepackage{hyperref}
\hypersetup{hidelinks}
\usepackage{microtype}
\usepackage{bm}

\newtheorem{proposition}{Proposition}
\newtheorem{theorem}{Theorem}
\newtheorem{corollary}{Corollary}
\newtheorem{lemma}{Lemma}

\newcommand{\cB}{c_{\rm best}}
\newcommand{\UB}{U}
\newcommand{\lift}{\mathcal{L}}

\begin{document}
\title{Iterative Quantum Optimization: From Greedy Reductions to Branch-and-Bound}
\author{Stuart Hadfield}
\email{shadfield@usra.edu} %
\affiliation{Quantum Artificial Intelligence Laboratory (QuAIL), NASA Ames Research Center, Moffett Field, CA 94035, USA}
\affiliation{USRA Research Institute for Advanced Computer Science, Mountain View, CA 94043, USA}
\author{Lucas T. Braydwood}
\affiliation{Quantum Artificial Intelligence Laboratory (QuAIL), NASA Ames Research Center, Moffett Field, CA 94035, USA}
\author{Eleanor Rieffel}
\thanks{Currently at the University of Queensland, Australia.}
\affiliation{Quantum Artificial Intelligence Laboratory (QuAIL), NASA Ames Research Center, Moffett Field, CA 94035, USA}
\date{\today}
\begin{abstract}
Iterative quantum optimization (IQO) uses quantum-generated information to select a sequence of classical problem reductions, dynamically exploring a single computational path that results in a feasible approximate solution. We introduce Bound-Aware Tree-IQO which combines this iterative structure with bounds and reversible search in the spirit of classical branch-and-bound algorithms. Feasible measured samples are used to update a shared incumbent, the best feasible solution found so far, while valid classical bounds prune computational branches that provably cannot improve it. Following one iteratively selected path recovers ordinary IQO. Incorporating backtracking yields an anytime method that heuristically returns the best feasible solution found when interrupted, or can be run as a complete search over the pruned tree which is guaranteed to return and certify the exact optimum when no unexplored nodes remain. For suitable single-variable or pairwise reductions, an initial IQO path can determine the branching schedule without further quantum calls required, though certifying the optimum may take superpolynomial classical time. To motivate our framework, we prove two limitations of irreversible IQO reductions based on the quantum approximate optimization algorithm (QAOA). We analytically show an infinite family of unweighted MaxCut instances on which globally optimized level-1 QAOA makes standard Recursive QAOA strictly prefer a first problem reduction choice that excludes every optimal solution, and limits that branch to an approximation ratio of at most $2/3+o(1)$. In contrast, a single backtrack to the complementary branch recovers an optimal solution using the same quantum information. Causal cone arguments give a complementary obstruction at every fixed depth of standard QAOA: a weighted MaxCut instance is constructed and shown to contain two edges with identical two-qubit reduced states (and hence identical local quantum guidance), even though every optimum requires opposite edge parities. MaxCut experiments show that QAOA-guided ordering is nonrandom, with a signal persisting in exact-certification tests through $n = 28$, while demonstrating that quantum samples can strengthen the incumbent and enable further pruning. Our results suggest that quantum data could provide useful guidance toward solving hard combinatorial optimization problems more effectively than either quantum or classical algorithms alone.
\end{abstract}

\maketitle
\begin{figure*}[t!]
\centering
\includegraphics[width=\textwidth]{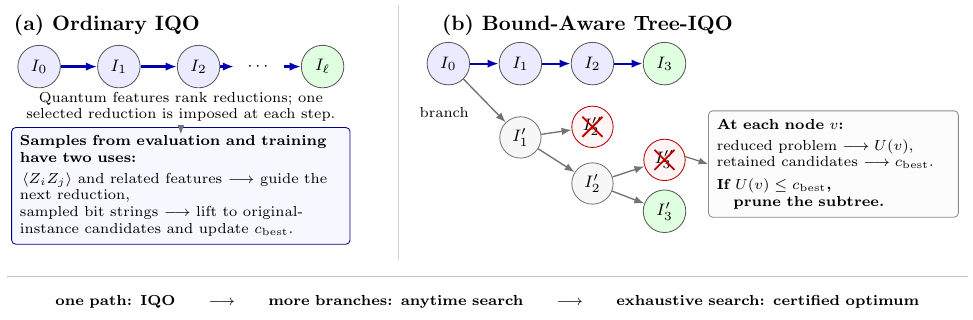}
\caption{From ordinary IQO to Bound-Aware Tree-IQO. Nodes denote occurrences of reduced problem instances along reduction histories and arrows denote reductions. Equivalent states may be merged in an implementation. (a) Ordinary IQO follows a single preferred sequence. Computational basis samples generated while evaluating or training the quantum subroutine can serve two roles: they estimate reduction features and, after lifting, provide original-instance candidates that update the primal bound $c_{\rm best}$. Any improvement raises the pruning threshold globally. (b) Bound-Aware Tree-IQO prunes any node whose valid upper bound satisfies $U(v)\le c_{\rm best}$, so one useful measured candidate can make an entire descendant subtree unnecessary. Tree-IQO retains the complementary outcomes of earlier reductions as alternative branches, so backtracking can revisit choices made along the initial IQO path. Throughout the paper we use a maximization convention, so $c_{\rm best}$ is a feasible lower bound and $U(v)$ is an upper bound. One preferred path recovers ordinary IQO. Broader exploration gives an anytime method, and exhaustive exploration of all unpruned branches certifies the optimum.}
\label{fig:framework_overview}
\end{figure*}

\section{Introduction}
Achieving a practical quantum advantage for hard combinatorial optimization remains an elusive goal~\cite{Abbas2024}.
Fault-tolerant processors may permit deeper optimization subroutines, while near-term devices impose tighter circuit budgets.
In either setting, a quantum subroutine need not solve the full problem instance on its own, and can instead be used for more limited but important tasks within a hybrid setting. Iterative quantum optimization (IQO) algorithms take this approach, using quantum-generated information to guide a sequence of problem reductions, while classical computation controls the overall procedure.
A greedy IQO run follows a single reduction path, committing to each selected outcome without backtracking.
The quantum approximate optimization algorithm (QAOA)~\cite{Farhi2014QAOA,Hadfield2019QAOA} underlies Recursive QAOA (RQAOA), a canonical early example.
Bravyi \emph{et al.}~\cite{Bravyi2020RQAOA} use pairwise QAOA correlations to choose a variable relation to fix, substitute it into the cost Hamiltonian, and iterate on the subsequently reduced problems until simplified enough to solve exactly.
See Appendix~\ref{app:maxcut} for the pairwise MaxCut reduction rules.
Later work has added problem-specific reductions, logical inference and hard constraint handling, learned recursive policies, and compressed relaxations, as well as application-specific variants~\cite{Dupont2023,BradyHadfield2024,Patel2024,Finzgar2024,Kondo2025,Mooney2026,Bae2025,RamosRuiz2026}.
Recent IQO algorithms for maximum independent set push the same idea further by combining shallow quantum information with large classical reductions or kernelization~\cite{BradyHadfield2024,WyboGreedy2026,Schuetz2025qReduMIS,Yalovetzky2026Portfolio}.
Our starting point is that IQO should make full use of the classical side of this division of labor. Reduction, inference, bounding, and search are mature classical tools, and many can be imported directly instead of being rebuilt inside the quantum routine.
Quantum-Informed Recursive Optimization (QIRO)'s problem-specific reductions and backtracking along its initial reduction path~\cite{Finzgar2024}, together with our complete tree search Quantum Davis--Putnam--Logemann--Loveland (DPLL) algorithm for Boolean satisfiability (SAT)~\cite{BradyHadfield2024,BradyHadfield2026}, already point in this direction.
Here, we propose Bound-Aware Tree-IQO (Fig.~\ref{fig:framework_overview}), which searches alternative quantum-guided reduction paths beyond the initially selected path, using objective bounds to prune computational branches. It returns the best feasible solution found when interrupted, or finds and certifies an optimum when run to completion.

Several observations make our extension especially natural. IQO reductions can be designed such that each feasible computational basis sample from any reduced problem directly lifts to a candidate solution for the original problem, with the same objective value. This is especially useful as many applicable quantum subroutines, including but not limited to QAOA, may produce many measured samples during parameter training and expectation value estimation.
The classical search controller keeps track of the best candidate found at any step in the tree search. Its value ($c_{\mathrm{best}}$) is the \emph{shared incumbent}.
A valid upper bound on the reduced problem also bounds the objective value achievable by every solution below that node (i.e., resulting from further reduction steps). Together they give the usual branch-and-bound pruning test.
Keeping both outcomes of selected reductions gives \emph{Bound-Aware Tree-IQO}: the preferred path is ordinary IQO. Exploring additional branches can improve the initial solution, while retaining the best solution found.
The key change is logical rather than quantum: a preferred relation may guide search without being promoted to a permanent reduction.
We elucidate this distinction by constructing an infinite family of unweighted MaxCut instances for which globally optimized level-$1$ QAOA makes standard RQAOA strictly prefer a first relation that excludes every optimum, limiting that branch to an approximation ratio of at most $2/3+o(1)$.
The complementary root branch contains an optimal solution and can be solved after one backtrack, with no new root quantum information.
For every fixed QAOA depth, we also construct a weighted MaxCut instance containing two edges whose reduced density matrices are identical for every choice of QAOA angles, even though every optimal cut requires the two edges to have opposite parities.
Thus no rule based only on the local state can fix the correct parity on both edges.
In both examples the quantum output can still be useful as a preference to guide which branches to explore first.

\paragraph*{Main results.} We show that every feasible sample at a reduced problem node lifts to a candidate solution for the original instance, and that valid node bounds safely prune paths that cannot improve the current incumbent (Propositions~1 and~2). When run to complete tree coverage, Bound-Aware Tree-IQO thus both finds and certifies the optimum (Theorem~1). For a selected set of pairwise parity fixing reductions (as in RQAOA) that form a spanning tree, the initial IQO root-to-leaf path can supply a complete fixed branching schedule, and so the remaining search requires no further quantum calls (Proposition~3 and Corollary~1).

We then give two separations showing that Bound-Aware Tree-IQO overcomes a fundamental limitation of prior IQO and RQAOA approaches. Theorem~2 gives an unweighted MaxCut family on which globally optimized level-1 RQAOA strictly prefers a first reduction that excludes every optimal solution. The complementary branch recovers an optimum after a single backtrack. Proposition~\ref{prop:fixed-depth-ambiguity}~in Appendix~\ref{app:greedytrap} shows a different obstruction that extends to every fixed QAOA depth; two edges can have matching correlator expectation values despite requiring opposite parities in any optimal solution.

We numerically test the Bound-Aware Tree-IQO algorithm on weighted MaxCut instances. QAOA correlations give a branching order that visits fewer nodes than random ordering on the tested ensembles, although strong classical ordering remains competitive. The benchmarked problem instances are small enough to verify exactly. The appendices contain additional results including stronger classical branching and incumbent control tests, and additional shot count, bound, and sparse-graph experiments. These tests separate the effects of branch order and sample reuse. They do not establish an overall runtime advantage alone. In practice, reusing feasible samples obtained throughout the IQO run, including during training, can improve the incumbent and reduce subsequent search, with the overall runtime benefit depending on implementation details.

The remainder of the paper is organized as follows. Section~\ref{sec:framework} introduces the generalized IQO framework, including problem reductions, sample reuse, and classical node bounds. Section~\ref{sec:tree} develops the tree search extension and presents the MaxCut analytical separation and numerical benchmarks. Section~\ref{sec:related} elaborates on related work, with future directions discussed in Section~\ref{sec:discussion}. The appendices provide additional technical derivations and proofs, implementation details, and numerical tests.

\section{Bound-aware tree iterative quantum optimization}
\label{sec:framework}
To introduce Bound-Aware Tree-IQO more formally, we consider a maximization problem with classical objective $c_0(x)$ over a feasible set $\mathcal{X}_0$. For minimization, the bound directions below are reversed.
A reduced state $I_v$ contains a feasible set ($\mathcal X_v$), a classical
objective function~$c_v$, and a record of the reductions used to reach it.
This record defines a lifting map
($L_v:\mathcal X_v\to\mathcal X_0$): reversing the recorded reductions
turns an assignment ($y\in\mathcal X_v$) into a feasible assignment
($L_v(y)$) of the original problem. For the pairwise parity reductions
used below, the record specifies which spins were identified and
their relative signs. The state may also retain information needed
to compute a valid node bound.
An IQO run produces a sequence of such reduced problems
\begin{equation}
I_0 \longrightarrow I_1 \longrightarrow \cdots \longrightarrow I_\ell,
\label{eq:path}
\end{equation}
where each edge corresponds to a selected problem reduction.
At each node in the sequence we also have a diagonal Hamiltonian~\cite{lucas2014ising,Hadfield2021Hamiltonians} representing the current classical objective function which is the input in the standard QAOA realization. Other QAOA variants or more general quantum subroutines may use a different encoding.
A reduction history fixes relations among original variables.
We define the reduced objective so that every feasible assignment has
the same value as its lifted original assignment
\begin{equation}
c_v(y)=c_0\!\left(\lift_v(y)\right).
\label{eq:costpreserve}
\end{equation}
Thus a feasible solution of a reduced problem immediately supplies a candidate of known and identical objective value for the original instance.
At a reduced state $v$, a quantum subroutine $Q(I_v)$ can provide two kinds of information: scores that rank reductions and sampled assignments that may already be feasible candidates. RQAOA uses pair correlations $\langle Z_iZ_j\rangle$ for the first task, while computational basis measurements can also serve the second. Quantum and classical information may be combined when choosing the next relation. A one-path IQO run imposes one outcome. A complete search retains enough outcomes to cover the current feasible set.
Each outcome defines a child problem. With hard constraints in particular, the assumed outcome is incorporated into the objective and constraints of the child, and sound logical inference may simplify it further, as in the constrained IQO and Quantum DPLL frameworks~\cite{BradyHadfield2024,BradyHadfield2026}. %
The proofs below require only correct reduction, lifting, bounds, and coverage. QAOA is not required for correctness: quantum information ranks branches or supplies candidates, while pruning is justified independently.

\subsection{Measured shots as primal bounds}
Suppose $Q(I_v)$ is measured in a basis that returns a classical candidate $y\in\mathcal{X}_v$. For unconstrained binary optimization in the computational basis, every such shot has this property.
Measurements in other bases need not produce candidate strings, and hard constraints can make a computational basis string infeasible. If decoding or repair yields a feasible reduced assignment ($\tilde y\in\mathcal X_v$), we evaluate $c_v(\tilde y)$ and lift $L_v(\tilde y)$. The raw infeasible string does not itself supply the incumbent.

\begin{proposition}[Sample lifting and incumbents]
\label{prop:incumbent}
For any node $v$ and any feasible sampled solution $y\in\mathcal{X}_v$, $x=\lift_v(y)$ is a feasible solution of the original problem with objective value $c_0(x)=c_v(y)$. Consequently
\begin{equation}
\cB=\max_{(v,y)\in\mathcal{S}} c_v(y)
\label{eq:incumbent}
\end{equation}
is a valid running lower bound on the optimum, where $\mathcal{S}$ contains the feasible samples retained so far.
\end{proposition}
Equation~\eqref{eq:costpreserve} establishes the proposition. In the maximization convention of this paper, $\cB$ is a \emph{primal lower bound}, witnessed by an explicit feasible solution. It should not be confused with the upper bounds of Eq.~\eqref{eq:ubdef}, which may come from relaxations, spectral arguments, or other bounding routines and need not themselves provide a feasible assignment.
Whenever a feasible shot raises $\cB$, it strengthens pruning throughout the open search: one improved candidate may eliminate many open search nodes and everything below them. In fixed-order Tree-IQO this saves classical search because the quantum schedule has already been generated. In an adaptive version it can also remove states before another quantum call is made. The wall-clock value is implementation dependent, but using a string that has already been measured requires no additional quantum sampling.
For MaxCut, the computational basis shots used to estimate $\langle Z_iZ_j\rangle$ are also candidate cuts. Sample-based training can produce many more such strings. With $N_{\rm eval}$ objective evaluations and $S$ shots, there are up to $N_{\rm eval}S$ candidates at that node, including samples from intermediate parameter values. If those strings are retained, no separate candidate-sampling stage is needed. The state preparations, optimization, and shots remain part of the quantum cost. The extra work is classical lifting, evaluation, and storage. The benefit is simply that one measurement record can be used twice: first to guide a reduction and then, when useful, to improve the incumbent. Measurement allocation can therefore matter in practice~\cite{LeeKim2026}. A poor sample is harmless because the incumbent changes only when a better feasible solution is found, and classical heuristics or exact subroutines can contribute candidates in exactly the same way.

\paragraph{Feasibility-preserving quantum subroutines.}
Our framework is naturally compatible with quantum algorithms engineered to output only feasible solutions.
Let $\Pi_v=\sum_{y\in\mathcal X_v}|y\rangle\langle y|$ project onto
the feasible computational basis subspace at node $v$. If the initial
state satisfies $\rho_v=\Pi_v\rho_v\Pi_v$ and every subsequent unitary
commutes with $\Pi_v$, ideal computational basis samples lie in
$\mathcal X_v$ with probability one. For QAOA, constraint-preserving mixers
provide this option when a suitable feasible initialization is
available~\cite{Hadfield2019QAOA}. Each sample then lifts directly to
a witnessed primal bound, without requiring repair or further postprocessing. This feasibility guarantee certifies neither
optimality nor a preferred reduction. Pruning still requires a valid
node bound.

\subsection{Node bounds and pruning}
Let $\UB(v)$ be any efficiently computable upper bound satisfying
\begin{equation}
\max_{y\in\mathcal{X}_v}c_v(y)\le \UB(v).
\label{eq:ubdef}
\end{equation}
Generic coefficient-norm bounds are always available for diagonal Hamiltonians, while problem-specific graph, spectral, semidefinite, or relaxation bounds can be much stronger. The generic bound is cheap but can be loose in practice. The search framework does not depend on how $\UB$ is obtained.

\begin{proposition}[Subtree pruning]
\label{prop:prune}
If $\UB(v)\le \cB$, then no descendant of node $v$ can improve the incumbent. The node and its entire descendant subtree may be discarded without changing the best solution returned by the search.
\end{proposition}
\begin{proof}
Every descendant corresponds to adding further restrictions to $\mathcal{X}_v$, so its feasible solutions lift through $v$ and have objective value at most $\max_{y\in\mathcal{X}_v}c_v(y)\le \UB(v)\le\cB$.
\end{proof}
The criterion $\UB(v)\le\cB$ is appropriate when the goal is the optimal objective value or one optimal witness. If the goal is to enumerate \emph{all} optimal solutions, equality branches must instead be retained and only $\UB(v)<\cB$ may be pruned.
The same test can stop an ordinary one-path IQO run: if a valid bound on the current reduced problem cannot beat the best feasible sample already found, continuing that path cannot help. With backtracking, the same pair of bounds becomes the basic pruning rule for Tree-IQO.
\section{From an IQO path to tree search}
\label{sec:tree}
Ordinary IQO follows a single reduction path, Eq.~\eqref{eq:path}. For a binary relation $O_t=\sigma_t$, retaining the complementary outcome $O_t=-\sigma_t$ creates an alternative child. Labeling each node by its full ordered reduction history gives a binary \emph{history tree}, which is the representation used in the proofs and numerics. Different histories can still reduce to equivalent states. An implementation may canonicalize and merge such states, giving a quotient directed acyclic graph (DAG). We keep the name Tree-IQO for the underlying history-tree construction.
This construction builds directly on Quantum DPLL's complete SAT backtracking~\cite{BradyHadfield2026}. Optimization adds objective information absent from pure feasibility search. Feasible solutions define an incumbent, and valid node bounds prune branches that cannot improve it. Backtracking is the general mechanism for revisiting an earlier choice; branch-and-bound is a backtracking tree search equipped with an incumbent and valid node bounds that certify which subtrees cannot improve it. In the near-term regime, shallow quantum output serves as heuristic branch guidance. Backtracking keeps those preferences reversible.
A particularly simple realization, and the one we study numerically below, uses the first IQO run to determine an \emph{ordered set of branching relations}. We call this \emph{fixed-order Tree-IQO}: the quantum subroutine learns the relation schedule once on the initial path, and the later certifying traversal reuses that schedule without rerunning the quantum subroutine at every branch. The subsequent search holds this schedule fixed and classically explores the preferred parity and its opposite at each level. The stored predicates may be viewed as relations on the original variables and are rewritten through the substitutions accumulated on each branch. Predicates already implied by earlier choices are simply skipped. For pairwise MaxCut reductions, $n-1$ independent parity relations form a spanning tree in the relation graph over the original spins and specify a complete cut assignment up to the global $\mathbb{Z}_2$ symmetry. More generally, a branch predicate may have more than two exhaustive outcomes, and one may stop at a cutoff size $\ell_0$ or whenever the residual instance enters a class that is efficiently solvable.
A general bound-aware realization needs five logical components:
\begin{enumerate}[leftmargin=*,nosep]
\item a branching rule that selects a relation or predicate and orders its retained outcomes, using quantum or classical information;
\item correctness-preserving reduction, lifting, and any sound logical inference needed to construct each child state;
\item a shared incumbent $\cB$ from explicit feasible candidates, including useful quantum samples;
\item a valid upper bound $\UB(v)$ at every reduced problem node; and
\item a search controller that selects open search nodes, detects equivalent states when desired, and enforces the chosen time and memory budget.
\end{enumerate}
The fixed-order implementation specializes item 1 to the schedule learned on the initial IQO line. The ordering rule itself may be quantum or classical. Exactness depends on exhaustive branch coverage and safe pruning, not on how good that ordering is. A depth-first implementation stores only the current root-to-node stack of reduced states, apart from any optional duplicate table. %

\begin{proposition}[Coverage of a fixed pairwise schedule]
\label{prop:pairwise-schedule}
Suppose an IQO path on $n$ binary spins performs $n-1$ pairwise contractions, each joining two previously disjoint components of the accumulated relation graph. Record the corresponding relations between original spins. The recorded pairs form a spanning tree. Branching exhaustively on both parities of those $n-1$ relations gives a bijection between leaves and spin assignments modulo the global spin flip. Consequently the fixed schedule covers every MaxCut assignment and can be traversed without refreshing the quantum subroutine.
\end{proposition}
\begin{proof}
Initially every spin is its own relation component. Each contraction joins two components and reduces their number by one, so $n-1$ contractions produce a connected acyclic relation graph, hence a spanning tree. Fix one root spin to $+1$. A parity choice on every tree edge then determines all other spins uniquely by propagation. Conversely, every spin assignment determines a unique vector of tree-edge parities. Removing the arbitrary root sign identifies precisely the global $\mathbb{Z}_2$ pair.
\end{proof}
For more general IQO reductions, fixed-order certification is available only when an analogous transport and coverage condition is verified. Otherwise Tree-IQO must refresh its branch generator or use a classical exhaustive predicate family at the affected node.
\subsection{Anytime and exact modes}
If the search is stopped after any fixed node or wall-clock budget, it returns the best incumbent found so far. Since the initial IQO solution is retained, its objective value can never be worse than that of the original one-path run, and exploring more nodes can only improve $\cB$. If the retained open search nodes cover all as-yet-unexplored feasible solutions, then
\begin{equation}
\UB_{\rm global}=\max\!\left\{\cB,\max_{v\in\mathcal{F}}\UB(v)\right\}
\end{equation}
for the current frontier $\mathcal{F}$ (with the second maximum omitted when the frontier is empty) is a valid global upper bound. Thus an interrupted run can report a rigorous optimality gap in addition to a heuristic incumbent. When $0<\cB\le\UB_{\rm global}$, the ratio $\cB/\UB_{\rm global}$ is a certified approximation ratio.
If instead every unpruned branch is explored and every terminal residual problem is solved exactly, the same framework is complete and the gap closes to zero.
\begin{theorem}[Completeness of Tree-IQO]
\label{thm:complete}
Assume that the retained branching process terminates after finitely many reductions on every branch, that the descendant terminal states collectively cover the original feasible set, that $\UB(v)$ is valid at every node, and that every unpruned terminal state is solved exactly. Then exhaustive Tree-IQO returns a globally optimal solution and certifies its optimality.
\end{theorem}
\begin{proof}
Without pruning, coverage by the branch family ensures that every feasible original solution belongs to at least one terminal state and is therefore considered. By Proposition~\ref{prop:prune}, pruning removes only subtrees whose best possible objective is no larger than an already known feasible solution. Hence no subtree containing a strictly better solution is removed. When all remaining states have been exhausted, the incumbent is globally optimal. The absence of an unpruned node with $\UB(v)>\cB$ supplies the certificate.
\end{proof}
For the fixed-order realization above, the quantum computer is used only on the initial root-to-leaf path. The remaining tree traversal is classical, so the quantum-resource requirement is comparable to a single IQO run even though the worst-case classical search remains exponential.
\begin{corollary}[Fixed-order quantum budget]
\label{cor:fixed-quantum-budget}
Suppose an initial IQO run uses $Q$ quantum subroutine calls and $S$ total shots to determine a transportable fixed branching schedule satisfying the coverage condition of Theorem~\ref{thm:complete}, together with an incumbent. Exhaustive fixed-order Tree-IQO can subsequently certify the optimum with no further quantum calls or shots. For pairwise reductions that fix one independent relation per step and continue to one representative spin, $Q=n-1$ when one quantum subroutine invocation, including its parameter-optimization loop, is used per reduction. Recycling the $S$ samples as incumbent candidates therefore changes only the classical post-processing, while the additional search cost is entirely classical and may remain exponential.
\end{corollary}
Here the extra tree traversal is classical. This differs from coherent quantum branch-and-bound algorithms, which seek to speed up the traversal itself~\cite{Montanaro2020,Chakrabarti2022}. One could also rerun the quantum subroutine after selected backtracks, perhaps using more depth or shots on the smaller residual problems. Such refreshes are compatible with exact search as long as they only reorder an exhaustive branch family. We leave the resulting quantum-budget tradeoff to future work and use a fixed schedule in the experiments below.
\subsection{A provable separation between greedy and backtracking IQO}
\label{sec:greedytrap}
Retaining alternative branches can be qualitatively important. There are simple MaxCut instances for which idealized RQAOA makes a strictly preferred first reduction that excludes every globally optimal solution. A single backtrack can nevertheless recover the optimum.
For $m\geq3$, let $G_m$ be obtained from the complete bipartite graph $K_{3,m}$, with bipartition $\{u,v,w\}\cup\{b_1,\ldots,b_m\}$, by adding the two internal edges $(u,v)$ and $(v,w)$. We consider standard RQAOA at level $p=1$, with exact expectation values and globally optimized QAOA parameters at the root. Because the unweighted $p=1$ objective is periodic, global optimization may be taken over a full periodic domain, for example $\beta\in[0,\pi)$ and $\gamma\in[0,2\pi)$. The RQAOA rule selects an edge of maximum $|\langle Z_iZ_j\rangle|$ and fixes its parity to $\operatorname{sgn}\langle Z_iZ_j\rangle$.
For a possibly weighted graph $G$, let $\mathrm{OPT}(G)$ denote
the maximum total weight of a cut of $G$. We sometimes write
$\mathrm{OPT}(G\mid R)$ for the maximum over cuts satisfying
a specified parity relation~$R$.
\begin{theorem}[Greedy RQAOA trap with one-backtrack recovery]
\label{thm:greedy-rqaoa-trap}
For every $m\geq3$, $\operatorname{OPT}(G_m)=3m$ and every optimum satisfies $Z_uZ_v=Z_vZ_w=+1$. Moreover, there exists $m_0$ such that, for every $m\geq m_0$ and every globally optimal level-$1$ QAOA parameter choice, the two internal edges are the only edges attaining the largest correlation magnitude and satisfy
\begin{equation}
\langle Z_uZ_v\rangle=\langle Z_vZ_w\rangle<0.
\end{equation}
The equality of the two internal correlators holds at all parameters by the $u\leftrightarrow w$ automorphism. Hence, under either tie-break between them, the first standard RQAOA reduction fixes one of these two relations to the wrong sign. Every cut in that branch has value at most $2m+2$, so its best possible approximation ratio is at most
\begin{equation}
\frac{2m+2}{3m}=\frac23+\frac{2}{3m}\xrightarrow[m\to\infty]{}\frac23.
\label{eq:trap-ratio-main}
\end{equation}
The complementary root branch contains a global optimum and, after the corresponding contraction, can be solved exactly in $O(m)$ time. Thus one root backtrack recovers an exact solution for the entire family.
\end{theorem}
\begin{figure*}[t]
\centering
\includegraphics[width=0.97\textwidth]{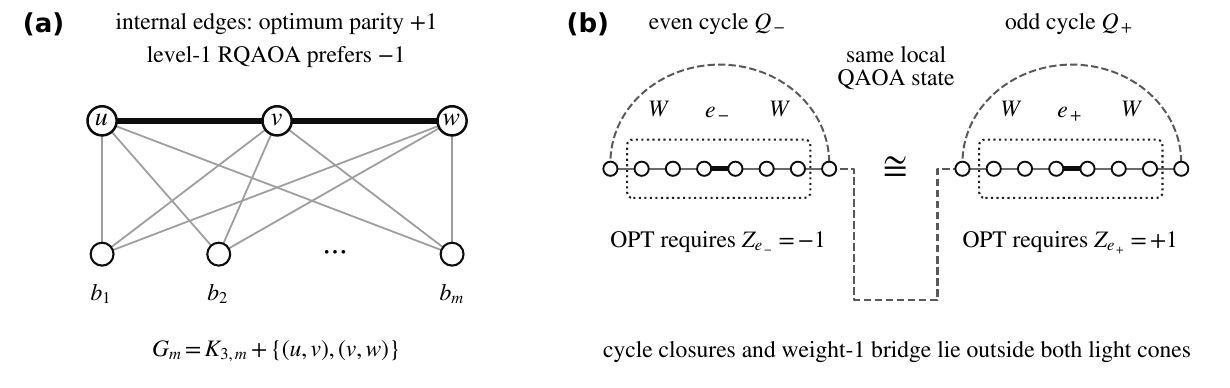}
\caption{The two separation constructions. (a) $G_m$ is $K_{3,m}$ plus the internal edges $(u,v)$ and $(v,w)$. Every optimum places $u,v,w$ on the same side of the cut, so both internal parities are $+1$, whereas globally optimized level-$1$ QAOA eventually makes these two edges the unique correlation-magnitude maximizers with negative sign. (b) For any fixed QAOA depth $p$, $H_{p,W}$ joins an even and an odd weighted cycle outside the depth-$p$ light cones of distinguished edges $e_-$ and $e_+$. Their rooted weighted depth-$p$ neighborhoods are identical, hence their two-qubit QAOA reduced states agree up to endpoint relabeling for every choice of the standard graph-global layer angles, while every optimum requires opposite parities. Dashed portions lie outside the indicated local neighborhoods; the schematic suppresses the number of weight-$W$ edges along each arm.}
\label{fig:theory-constructions}
\end{figure*}
\begin{corollary}[Same-information separation]
\label{cor:same-information}
On the family $G_m$, the difference between the greedy and backtracking outcomes does not require a stronger root quantum state or more accurate root correlations. Standard one-path RQAOA irreversibly follows the strictly preferred branch and is confined to value at most $2m+2$, whereas a Tree-IQO policy using the same optimized level-$1$ root information and exploring the complementary root branch is exact after one backtrack.
\end{corollary}
The proof is in Appendix~\ref{app:greedytrap}. Patel \emph{et al.} previously gave a finite weighted example in which energy-optimal QAOA angles assign the wrong sign to the most strongly correlated edge~\cite{Patel2024}. Here the failure occurs on an infinite family of \emph{unweighted} instances, the standard all-edge RQAOA rule strictly selects the wrong edge, and the loss approaches a constant factor. The graphs are simple by design: the point is to expose the cost of making a shallow-state preference irreversible, not to propose a hard MaxCut benchmark. Any search that eventually tries both root parities escapes this trap.
The theorem concerns globally optimized root parameters. At finite size the optimizer has competing branches: at $m=13$, for example, the trap branch beats the benign branch by only $1.41\times10^{-3}$ in QAOA objective value. A local or finite-precision optimizer can therefore miss this particular failure. That does not change the separation, but it is a useful reminder that the output of a shallow variational calculation is a heuristic preference, not a logical certificate.
The RQAOA trap illustrates the broader distinction for heuristic IQO reductions. A one-path method that commits to a quantum-preferred relation must live with that choice. Tree-IQO only needs a ranking that tends to send the search toward useful branches early. A bad ranking can cost time, but it need not remove the optimum. On $G_m$, level-$1$ QAOA gives its largest correlation magnitude to the two internal relations and prefers the parity opposite to that required by every optimum. Tree-IQO keeps that preference as a branch-ordering heuristic instead of turning it into an irreversible constraint.
\paragraph{Fixed-depth local-state obstruction.} Proposition~\ref{prop:fixed-depth-ambiguity} in Appendix~\ref{app:greedytrap} goes beyond the correlator itself. For every fixed depth $p$, we construct a connected weighted instance with two edges whose \emph{entire two-qubit QAOA reduced states} agree up to endpoint relabeling for every choice of the standard graph-global layer angles, although every optimum requires opposite parities on the two edges. An isomorphism-equivariant rule that sees only this local quantum state therefore cannot be forced to choose a parity correctly on both. The construction uses the standard QAOA light (causal) cone~\cite{FarhiGamarnikGutmann2020,WurtzLove2021,HadfieldHoggRieffel2023}. It is weaker than Theorem~\ref{thm:greedy-rqaoa-trap} in one respect: the candidate pair is restricted and the two scores are tied. Whether the standard all-edge RQAOA rule has a strict trap of this kind for every fixed $p$ remains open.
In plain terms, the two edges look identical to every allowed local QAOA measurement, but the global optimum needs one edge cut and the other uncut. A mandatory rule using only that local record must therefore make the same decision on both and cannot be correct on both.
\subsection{MaxCut bound and numerical benchmark}
\label{sec:treenum}
For the shifted weighted MaxCut problem of Eq.~\eqref{eq:maxcutH}, a simple valid upper bound is
\begin{equation}
\UB(G;a)=a+\sum_{(ij):,w_{ij}>0}w_{ij},
\label{eq:maxcutub}
\end{equation}
obtained by allowing every positive-weight edge to be cut and every negative-weight edge to remain uncut independently. It is cheap and generally loose. Spectral, semidefinite, or problem-specific bounds can be substituted directly.
We first use a dense weighted MaxCut benchmark for which the full search can still be checked exactly. Each instance is a complete graph with i.i.d. edge weights uniform on $[0.2,1]$. At every IQO step we simulate depth-$1$ QAOA by exact state vectors and optimize its two angles numerically. The path continues to one remaining spin, leaving $n-1$ pair relations that define the fixed binary search tree. The rest of the traversal is classical and uses Eq.~\eqref{eq:maxcutub} for pruning. We use 50 independent instances at each $n\in\{10,12,14\}$. These are mechanism tests, not a solver-scale benchmark: the searched-node fraction measures tree structure and gives equal weight to node visits even when their classical processing costs differ.
For the classical branching baseline, we use differencing edge contraction (DEC)~\cite{HassinLeshenko2021}, which selects the current largest-weight edge and prefers opposite endpoint spins.
\begin{figure*}[t]
\centering
\includegraphics[width=\textwidth]{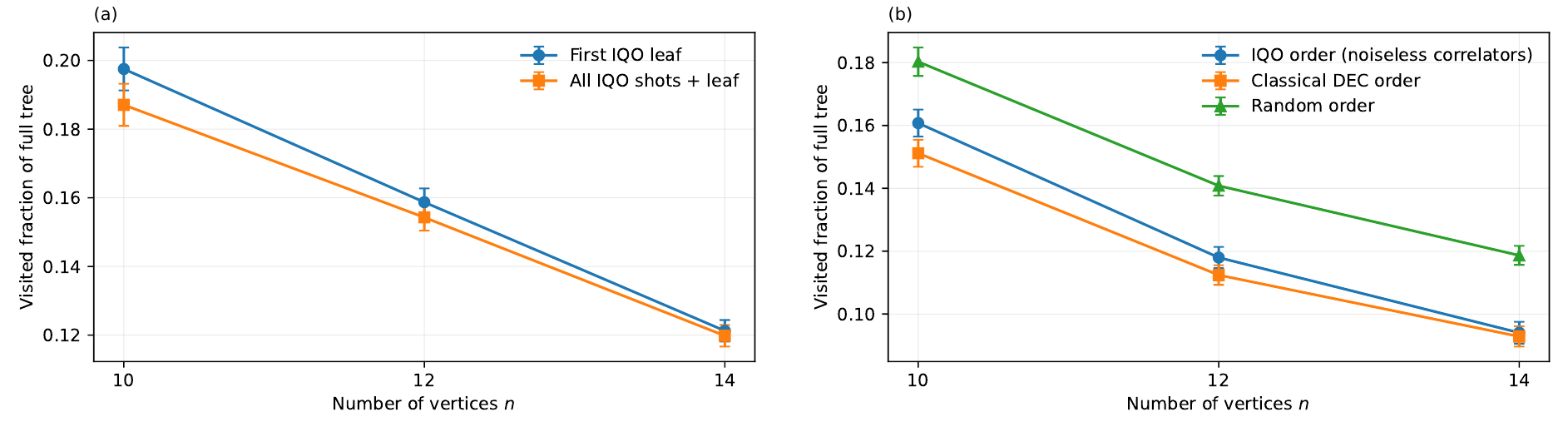}
\caption{Tree-IQO benchmark on complete weighted MaxCut, 50 instances per size. All quantum tests shown use only QAOA at level $p=1$. We expect deeper QAOA or suitable alternative ansatzes to improve branching guidance or candidate quality, although this remains to be tested. The vertical axis is the fraction of the full $2^n-1$ node binary tree visited before exact certification; error bars are one standard error. (a) Eight-shot batches both estimate the finite-shot IQO correlations and provide incumbent candidates. (b) With a common incumbent, branch ordering is determined by noiseless QAOA correlators at the numerically optimized angles, classical DEC, or random schedules; the random result averages three schedules per instance. All searches use Eq.~\eqref{eq:maxcutub}. Paired comparisons use the bootstrap procedure of Appendix~\ref{app:numerics}.}
\label{fig:treeprune}
\end{figure*}
The full history trees at $n=10,12,14$ contain $1023$, $4095$, and $16383$ nodes. Modern exact MaxCut solvers use much stronger relaxations and operate in a different regime~\cite{RendlRinaldiWiegele2010,KrislockMalickRoupin2017}.
Figure~\ref{fig:treeprune}(a) tests sample reuse. The best intermediate sample beats the first IQO leaf in $68\%$, $80\%$, and $78\%$ of instances, raising the mean initial approximation ratio from $0.959$, $0.952$, and $0.960$ to $0.982$, $0.976$, and $0.979$. The corresponding searched-tree fraction changes from $19.7\%$, $15.9\%$, and $12.1\%$ to $18.7\%$, $15.4\%$, and $12.0\%$. The effect on search is modest at these sizes, but it directly shows that measurements already used for IQO guidance can also strengthen the incumbent.
Figure~\ref{fig:treeprune}(b) isolates branch ordering. With a common incumbent, the noiseless-correlator IQO order visits $16.1\%$, $11.8\%$, and $9.4\%$ of the tree on average, compared with $15.1\%$, $11.2\%$, and $9.3\%$ for the classical DEC order and $18.0\%$, $14.1\%$, and $11.9\%$ for random ordering. The QAOA order is therefore clearly nonrandom on this ensemble, but it does not outperform DEC at the smaller two sizes. Appendix~\ref{app:analytic-ordering} gives a larger independent ordering-only study through $n=28$. In that expanded study the paired IQO-minus-random and IQO-minus-DEC intervals are negative at every tested size, with the smaller $n=26$ and $28$ cohorts retained as scale probes. This is evidence for an ensemble-specific ordering signal, not an end-to-end runtime advantage.
The two outputs of a quantum call play different roles. A better feasible sample can only improve the shared incumbent, whereas a different branch order may help or hurt. Appendix~\ref{app:extra_numerics} compares the QAOA order with a one-step child-bound brancher, adds QAOA samples on top of classical local-search incumbents, and checks the root angle search more aggressively. The qualitative picture is unchanged.
All quantum benchmarks here use $p=1$, where the two mechanisms are easiest to separate and the correlators are also easy to evaluate classically. Deeper circuits can change both the sampled candidates and the reduction scores. Proposition~\ref{prop:fixed-depth-ambiguity} shows, however, that no fixed depth turns local parity information into a universal certificate. Further finite-shot, bound, sparse-graph, and classical-reference tests are collected in Appendix~\ref{app:extra_numerics}.

\section{Related work}
\label{sec:related}
RQAOA first used QAOA correlations to select variable relations and %
iteratively eliminate variables~\cite{Bravyi2020RQAOA}.
A 2023 presentation~\cite{HadfieldAPS2023} %
described a branch-and-bound variant of IQO that could be applied heuristically or exhaustively, as well as different iterative reductions and hard-constraint extensions. The current paper now fully describes that approach.
Several ways to connect quantum information to classical reductions have subsequently been developed.
Constrained IQO incorporates problem-specific inference~\cite{BradyHadfield2024,BradyHadfield2026}, QIRO combines such reductions with heuristic backtracking~\cite{Finzgar2024}, and learned policies can change how recursive decisions are selected~\cite{Patel2024}. Other %
iterative variants
broaden the available reductions and quantum subroutines~\cite{Dupont2023,Kondo2025,WyboGreedy2026, Schuetz2025qReduMIS,Yalovetzky2026Portfolio}, including recent extensions to constraint-preserving QAOA mixers~\cite{RamosRuiz2026,bassa2026iterative}.
These techniques can each be similarly adapted to a Tree-IQO framework for a variety of optimization problems.

Our prior work on Quantum DPLL gives iterative guidance within complete %
tree search for the Satisfiability decision problem~\cite{BradyHadfield2026}.
Here we retain complementary outcomes of IQO reductions as branches of an optimization search. In Tree-IQO all feasible solution samples provide a shared objective incumbent, which together with valid upper bounds justifies pruning of computational branches. Exhaustive search certifies an optimum.

Complete search in which a classical controller is informed by quantum samples has %
precedent.
Tran \emph{et al.} use annealer samples from relaxed scheduling problems
within a classical search that enforces constraints and uses the samples
to guide exploration~\cite{Tran2016,TranMars2016}. Sanavio \emph{et al.}
delegate reduced integer-programming subproblems to an
annealer~\cite{Sanavio2024}, and Peng \emph{et al.} use Ising-solver output
for both branching guidance and candidate solutions inside classical
branch-and-bound~\cite{Peng2025}. Other related work includes
annealing-assisted bounding for transportation scheduling~\cite{Bozejko2026},
hybrid branch-and-bound for binary linear problems~\cite{Czegel2025QCBB},
and quantum-assisted column generation in branch-and-price for vertex
coloring~\cite{Vercellino2025QCBP}. H{\"a}ner \emph{et al.}'s
quantum-amenable QUBO solver is a classical branch-and-bound design
intended to support quantum search, rather than a method guided by
measured quantum output~\cite{HanerBooth2024}. 

Tree search has also been considered for quantum relaxations involving compression onto fewer qubits~\cite{Huang2023QRBnB,Matsuyama2024QRBnB}. There, Pauli rounding constructs incumbent candidates from quantum expectation values. Exact relaxation optima give valid node bounds, whereas unvalidated variational energy estimates need not~\cite{Matsuyama2024QRBnB}.
Bound-Aware Tree-IQO instead organizes search around reversible IQO
reductions %
with each feasible computational basis sample giving a candidate solution and pruning justified by valid classical bounds.

While backtracking, quantum-guided complete search, and the use of sampled candidates are not new, our novel contributions include the explicit wrong-first-relation separation for globally optimized level-$1$ RQAOA, the fixed-depth local-state obstruction, and a precise account of when an IQO reduction schedule can be reused for complete search without further quantum calls. The experiments separately measure the effects of sampled incumbents and branch order. They do not establish an end-to-end runtime advantage over these earlier hybrid or purely classical methods, leaving that question open.

The numerical ingredients also have close classical counterparts. LP and SDP correlations can outperform shallow QAOA correlations in some MaxCut regimes~\cite{Fischer2024Shrinking}, and exact MaxCut solvers use much stronger relaxations than the ones used here~\cite{RendlRinaldiWiegele2010,KrislockMalickRoupin2017}. Reusing measured candidates is related to sample persistence, warm starts, relax-and-round, surrogate sampling, and qReduMIS~\cite{Karimi2017,EggerMarecekWoerner2021,Tate2023WarmStart,DupontSundar2024,Schuetz2025qReduMIS,WyboFinzgar2026QISS}. Appendix~\ref{app:extra_numerics} therefore includes matched classical branching and incumbent controls.

Quantum acceleration can also act on branch filtering or tree traversal. Fully coherent quantum walk and branch-and-bound algorithms
accelerate tree exploration~\cite{Montanaro2018,Montanaro2020,Chakrabarti2022}.
Booth \emph{et al.} developed quantum-accelerated filtering subroutines within classical constraint-programming search~\cite{Booth2020}, with subsequent work also considering hybrid and fully quantum search schemes~\cite{Booth2021}. Hardware resource assumptions may determine design choices and
the position of the quantum subroutine %
within the overall algorithm.
\section{Discussion}
\label{sec:discussion}
Bound-Aware Tree-IQO combines quantum-guided problem reductions with classical search that allows revisiting earlier choices. Quantum information can contribute in two important ways, by choosing which branches to explore in which order, and by supplying feasible candidates that can strengthen the incumbent and reduce the number of nodes explored through pruning. Our separation results show our tree search extensions of IQO to be well-motivated. Globally optimized QAOA can select a problem reduction that excludes every optimal solution, which is only recovered from by backtracking to the complementary branch. In the fixed-order setting, a suitable initial IQO run can supply the branching schedule and candidate pool for a subsequent classical search, with no further quantum calls necessary for certification. In practice, additional IQO runs or classical heuristics could be employed to seek better incumbents.

The MaxCut experiments separately demonstrate these two contributions. QAOA-guided ordering contains a signal relative to random ordering on the tested ensembles, and reusing measured candidates can improve the incumbent and pruning. The comparisons also show why classical controls are essential. The classical greedy DEC algorithm often matches or improves on the tested QAOA ordering, and inexpensive classical local search supplies strong incumbents to begin with. All our quantum tests use only level-1 QAOA, whose MaxCut correlations are efficiently computable classically. These experiments therefore test search mechanisms and sample reuse, without establishing an overall runtime advantage. We expect suitable deeper or alternative ansatzes to improve guidance or candidate quality, but their additional costs must be included in that comparison.

The Tree-IQO framework applies to other discrete optimization problems when reductions preserve feasibility and objective value under lifting, retained branches cover the feasible set, valid bounds are efficiently obtainable, and terminal residual problems are solved exactly. Weighted MaxSAT and maximum independent set are natural settings with problem-specific reduction rules and inference. For constrained problems, measured strings can update the incumbent only when feasibility is established, directly through the quantum preparation or through appropriate decoding or repair. The MaxCut separation constructions and spanning-tree schedule do not automatically extend to these settings without adaptation.

Generally, practical usefulness depends on total computational cost resulting from various implementation details. Resource comparisons must include state preparation, parameter training, sampling, communication, and classical processing, as well as the search nodes saved. Completeness guarantees an optimum when search finishes, but gives no polynomial runtime guarantee. The difficulty of computing quantum scoring information must also be assessed at the precision relevant to search. Exact or exponentially precise MaxCut correlator evaluation on unrestricted graphs is computationally hard for every fixed QAOA level $p\geq2$~\cite{Wang2025ExpectationComplexity,Hadfield2026ExpectationComplexity}. Those results leave constant or inverse-polynomial additive error open, while fixed-depth correlators on bounded-degree graphs remain classically tractable. With backtracking, search need not perfectly resolve every near tie. An interesting direction is to better characterize tradeoffs between the number of measurement shots at each node and the size of the resulting search tree.

A useful next step is to compare deeper QAOA circuits along with other quantum ans\"atze on matched dense and sparse ensembles against strong classical branching rules and incumbent heuristics. Such tests should report time to a specified solution quality, certified gap versus time, and time to prove optimality, including the full training and sampling cost, facilitating comparison with a wealth of classical approaches. Hardware experiments can then assess how noise affects branching scores and incumbent candidates, and the tradeoffs between approaches to managing that noise. Adaptive Tree-IQO variants introduce the further question of when a fresh quantum call after backtracking is worth its cost. On the theory side, it remains open whether the strict all-edge trap extends to every fixed QAOA depth, and whether analogous obstructions occur for other problems and reduction rules. Ultimately, we seek to further develop hybrid quantum-
classical algorithms and better characterize when they are more effective than either component alone.

\begin{acknowledgments}
We thank our colleagues from NASA QuAIL for numerous helpful discussions.
We thank John Tanner at the University of Queensland for helpful comments on an early draft.
This material is based upon work supported by the U.S. Department of Energy, Office of Science, Office of Advanced Scientific Computing Research under Award No.~89243024SSC000129 and funded through interagency agreement SAA2-403857. S.H. was supported under Prime Contract No.~80ARC020D0010 with the NASA Ames Research Center.
OpenAI ChatGPT (GPT-5.6 Sol) was used for literature synthesis, manuscript reorganization and editing, and for improving the initial author developed code and numerical analysis from several years of draft versions of this manuscript. The authors %
have verified all claims and
take full responsibility for all content of the manuscript.
\end{acknowledgments}

\section*{Data and code availability}
The numerical analyses used recorded instance streams and fixed
pseudorandom seeds. The authors have retained source code,
analysis scripts, and instance-level outputs underlying the reported
results. %
The distribution of these materials is subject to institutional review and approval.
Inquiries %
may be directed to the corresponding author.
\bibliographystyle{source/quantum}
\bibliography{bib}

@article{Abbas2024,
  author = {Abbas, Amira and Ambainis, Andris and Augustino, Brandon and B{\"a}rtschi, Andreas and Buhrman, Harry and Coffrin, Carleton and Cortiana, Giorgio and Dunjko, Vedran and Egger, Daniel J. and Elmegreen, Bruce G. and Franco, Nicola and Fratini, Filippo and Fuller, Bryce and Gacon, Julien and Gonciulea, Constantin and Gribling, Sander and Gupta, Swati and Hadfield, Stuart and Heese, Raoul and Kircher, Gerhard and Kleinert, Thomas and Koch, Thorsten and Korpas, Georgios and Lenk, Steve and Mare{\v{c}}ek, Jakub and Markov, Vanio and Mazzola, Guglielmo and Mensa, Stefano and Mohseni, Naeimeh and Nannicini, Giacomo and O'Meara, Corey and {Pe{\~n}a Tapia}, Elena and Pokutta, Sebastian and Proissl, Manuel and Rebentrost, Patrick and Sahin, Emre and Symons, Benjamin C. B. and Tornow, Sabine and Valls, V{\'i}ctor and Woerner, Stefan and Wolf-Bauwens, Mira L. and Yard, Jon and Yarkoni, Sheir and Zechiel, Dirk and Zhuk, Sergiy and Zoufal, Christa},
  title = {Challenges and opportunities in quantum optimization},
  journal = {Nature Reviews Physics},
  volume = {6},
  number = {12},
  pages = {718--735},
  year = {2024},
  doi = {10.1038/s42254-024-00770-9}
}

@misc{Farhi2014QAOA,
  author = {Farhi, Edward and Goldstone, Jeffrey and Gutmann, Sam},
  title = {A quantum approximate optimization algorithm},
  year = {2014},
  eprint = {1411.4028},
  archivePrefix = {arXiv},
  primaryClass = {quant-ph},
}

@article{Hadfield2019QAOA,
  author = {Hadfield, Stuart and Wang, Zhihui and O'Gorman, Bryan and Rieffel, Eleanor G. and Venturelli, Davide and Biswas, Rupak},
  title = {From the quantum approximate optimization algorithm to a quantum alternating operator ansatz},
  journal = {Algorithms},
  volume = {12},
  number = {2},
  pages = {34},
  year = {2019},
  doi = {10.3390/a12020034}
}

@article{Bravyi2020RQAOA,
  author = {Bravyi, Sergey and Kliesch, Alexander and Koenig, Robert and Tang, Eugene},
  title = {Obstacles to variational quantum optimization from symmetry protection},
  journal = {Physical Review Letters},
  volume = {125},
  number = {26},
  pages = {260505},
  year = {2020},
  doi = {10.1103/PhysRevLett.125.260505}
}

@article{Dupont2023,
  author = {Dupont, Maxime and Evert, Bram and Hodson, Mark J. and Sundar, Bhuvanesh and Jeffrey, Stephen and Yamaguchi, Yuki and Feng, Dennis and Maciejewski, Filip B. and Hadfield, Stuart and Alam, M. Sohaib and Wang, Zhihui and Grabbe, Shon and Lott, P. Aaron and Rieffel, Eleanor G. and Venturelli, Davide and Reagor, Matthew J.},
  title = {Quantum-enhanced greedy combinatorial optimization solver},
  journal = {Science Advances},
  volume = {9},
  number = {45},
  pages = {eadi0487},
  year = {2023},
  doi = {10.1126/sciadv.adi0487}
}

@article{BradyHadfield2024,
  author = {Brady, Lucas T. and Hadfield, Stuart},
  title = {Iterative quantum algorithms for maximum independent set},
  journal = {Physical Review A},
  volume = {110},
  number = {5},
  pages = {052435},
  year = {2024},
  doi = {10.1103/PhysRevA.110.052435}
}

@article{Patel2024,
  author = {Patel, Yash J. and Jerbi, Sofiene and B{\"a}ck, Thomas and Dunjko, Vedran},
  title = {Reinforcement learning assisted recursive {QAOA}},
  journal = {EPJ Quantum Technology},
  volume = {11},
  number = {1},
  pages = {6},
  year = {2024},
  doi = {10.1140/epjqt/s40507-023-00214-w}
}

@article{Finzgar2024,
  author = {Fin\v{z}gar, Jernej Rudi and Kerschbaumer, Aron and Schuetz, Martin J. A. and Mendl, Christian B. and Katzgraber, Helmut G.},
  title = {Quantum-informed recursive optimization algorithms},
  journal = {PRX Quantum},
  volume = {5},
  number = {2},
  pages = {020327},
  year = {2024},
  doi = {10.1103/PRXQuantum.5.020327}
}

@article{Kondo2025,
  author = {Kondo, Ruho and Sato, Yuki and Raymond, Rudy and Yamamoto, Naoki},
  title = {Recursive quantum relaxation for combinatorial optimization problems},
  journal = {Quantum},
  volume = {9},
  pages = {1594},
  year = {2025},
  doi = {10.22331/q-2025-01-15-1594}
}

@article{Mooney2026,
  author = {Mooney, Gary J. and Villanueva, Jedwin and Bardhan, Bhaskar Roy and Ghosh, Joydip and Hill, Charles D. and Hollenberg, Lloyd C. L.},
  title = {Optimization-free recursive Quantum Approximate Optimization Algorithm for the binary paint shop problem},
  journal = {Physical Review Research},
  volume = {8},
  number = {1},
  pages = {013310},
  year = {2026},
  doi = {10.1103/xv79-w3rs}
}

@article{Bae2025,
  author = {Bae, Eunok and Kwon, Hyukjoon and Vijendran, V. and Lee, Soojoon},
  title = {Modified recursive {QAOA} for exact {Max-Cut} solutions on bipartite graphs: closing the gap beyond {QAOA} limit},
  journal = {Journal of Physics A: Mathematical and Theoretical},
  volume = {58},
  number = {48},
  pages = {485304},
  year = {2025},
  doi = {10.1088/1751-8121/ae20d7}
}

@misc{LeeKim2026,
  author = {Lee, Euimin and Kim, Shiho},
  title = {Adaptive shot allocation for recursive {QAOA} via reinforcement learning},
  year = {2026},
  eprint = {2605.26544},
  archivePrefix = {arXiv},
  primaryClass = {quant-ph},
}

@article{BradyHadfield2026,
  author = {Brady, Lucas T. and Hadfield, Stuart},
  title = {Quantum {Davis--Putnam--Logemann--Loveland} and generalized constraints in iterative quantum algorithms},
  journal = {Physical Review A},
  volume = {114},
  number = {2},
  pages = {022404},
  year = {2026},
  doi = {10.1103/8g3p-tyz1}
}

@article{HassinLeshenko2021,
  author = {Hassin, Refael and Leshenko, Nikita},
  title = {Greedy differencing edge-contraction heuristic for the {Max-Cut} problem},
  journal = {Operations Research Letters},
  volume = {49},
  number = {3},
  pages = {320--325},
  year = {2021},
  doi = {10.1016/j.orl.2021.02.006}
}

@article{Dunning2018,
  author = {Dunning, Iain and Gupta, Swati and Silberholz, John},
  title = {What works best when? {A} systematic evaluation of heuristics for {Max-Cut} and {QUBO}},
  journal = {INFORMS Journal on Computing},
  volume = {30},
  number = {3},
  pages = {608--624},
  year = {2018},
  doi = {10.1287/ijoc.2017.0798}
}

@article{Wang2018,
  author = {Wang, Zhihui and Hadfield, Stuart and Jiang, Zhang and Rieffel, Eleanor G.},
  title = {Quantum approximate optimization algorithm for {MaxCut}: A fermionic view},
  journal = {Physical Review A},
  volume = {97},
  number = {2},
  pages = {022304},
  year = {2018},
  doi = {10.1103/PhysRevA.97.022304}
}

@article{Ozaeta2022,
  author = {Ozaeta, Asier and van Dam, Wim and McMahon, Peter L.},
  title = {Expectation values from the single-layer quantum approximate optimization algorithm on {Ising} problems},
  journal = {Quantum Science and Technology},
  volume = {7},
  number = {4},
  pages = {045036},
  year = {2022},
  doi = {10.1088/2058-9565/ac9013}
}

@article{Karimi2017,
  author = {Karimi, Hamed and Rosenberg, Gili and Katzgraber, Helmut G.},
  title = {Effective optimization using sample persistence: A case study on quantum annealers and various {Monte Carlo} optimization methods},
  journal = {Physical Review E},
  volume = {96},
  number = {4},
  pages = {043312},
  year = {2017},
  doi = {10.1103/PhysRevE.96.043312}
}

@inproceedings{Tran2016,
  author = {Tran, Tony T. and Do, Minh and Rieffel, Eleanor G. and Frank, Jeremy and Wang, Zhihui and O'Gorman, Bryan and Venturelli, Davide and Beck, J. Christopher},
  title = {A hybrid quantum-classical approach to solving scheduling problems},
  booktitle = {Proceedings of the Ninth International Symposium on Combinatorial Search ({SoCS} 2016)},
  publisher = {{AAAI} Press},
  pages = {98--106},
  year = {2016},
  doi = {10.1609/socs.v7i1.18390}
}

@inproceedings{Peng2025,
  author = {Peng, Zedong and de Roux, Daniel and Bernal Neira, David E.},
  title = {Hybrid quantum branch-and-bound method for quadratic unconstrained binary optimization},
  booktitle = {Proceedings of Quantum Informatics, Computing \& Technology 2025 ({QC-HORIZON} 2025)},
  series = {Computer Science Research Notes},
  publisher = {V{\'a}clav Skala -- UNION Agency},
  year = {2025},
  editor = {Kais, Sabre and Skala, V{\'a}clav},
  volume = {3502},
  doi = {10.24132/CSRN.2025-A67}
}

@article{Bozejko2026,
  author = {Bo{\.z}ejko, Wojciech and Uchro\'nski, Mariusz and Wodecki, Mieczys{\l}aw},
  title = {Quantum annealing-driven branch and bound for the total weighted tardiness traveling salesman problem on the {{D-Wave}} quantum computer},
  journal = {International Transactions in Operational Research},
  volume = {33},
  number = {6},
  pages = {3855--3881},
  year = {2026},
  doi = {10.1111/itor.70161}
}

@article{Montanaro2020,
  author = {Montanaro, Ashley},
  title = {Quantum speedup of branch-and-bound algorithms},
  journal = {Physical Review Research},
  volume = {2},
  number = {1},
  pages = {013056},
  year = {2020},
  doi = {10.1103/PhysRevResearch.2.013056}
}

@article{Montanaro2018,
  author = {Montanaro, Ashley},
  title = {Quantum-walk speedup of backtracking algorithms},
  journal = {Theory of Computing},
  volume = {14},
  number = {15},
  eid = {15},
  pages = {1--24},
  year = {2018},
  doi = {10.4086/toc.2018.v014a015}
}

@inproceedings{Booth2020,
  author = {Booth, Kyle E. C. and O'Gorman, Bryan and Marshall, Jeffrey and Hadfield, Stuart and Rieffel, Eleanor G.},
  title = {{Quantum}-accelerated global constraint filtering},
  booktitle = {Principles and Practice of Constraint Programming},
  series = {Lecture Notes in Computer Science},
  volume = {12333},
  pages = {72--89},
  publisher = {Springer},
  year = {2020},
  doi = {10.1007/978-3-030-58475-7_5}
}

@article{Booth2021,
  author = {Booth, Kyle E. C. and O'Gorman, Bryan and Marshall, Jeffrey and Hadfield, Stuart and Rieffel, Eleanor G.},
  title = {Quantum-accelerated constraint programming},
  journal = {Quantum},
  volume = {5},
  pages = {550},
  year = {2021},
  doi = {10.22331/q-2021-09-28-550}
}

@misc{Chakrabarti2022,
  author = {Chakrabarti, Shouvanik and Minssen, Pierre and Yalovetzky, Romina and Pistoia, Marco},
  title = {Universal quantum speedup for branch-and-bound, branch-and-cut, and tree-search algorithms},
  year = {2022},
  eprint = {2210.03210},
  archivePrefix = {arXiv},
  primaryClass = {quant-ph},
}

@article{Sanavio2024,
  author = {Sanavio, Claudio and Tignone, Edoardo and Ercolessi, Elisa},
  title = {Hybrid classical--quantum branch-and-bound algorithm for solving integer linear problems},
  journal = {Entropy},
  volume = {26},
  number = {4},
  pages = {345},
  year = {2024},
  doi = {10.3390/e26040345}
}

@article{RamosRuiz2026,
  author = {Ramos-Ruiz, Pablo and Fuentes-Jim\'enez, Antonio Miguel and Ramos-Ruiz, Jos\'e E. and Jim\'enez-Manchado, Inmaculada},
  title = {On the use of the quantum alternating operator ansatz in quantum-informed recursive optimization: A case study on the minimum vertex cover},
  journal = {AppliedMath},
  volume = {6},
  number = {2},
  pages = {24},
  year = {2026},
  doi = {10.3390/appliedmath6020024}
}

@misc{FarhiGamarnikGutmann2020,
  author = {Farhi, Edward and Gamarnik, David and Gutmann, Sam},
  title = {The quantum approximate optimization algorithm needs to see the whole graph: Worst-case examples},
  year = {2020},
  eprint = {2005.08747},
  archivePrefix = {arXiv},
  primaryClass = {quant-ph},
}

@article{MoharPoljak1990,
  author = {Mohar, Bojan and Poljak, Svatopluk},
  title = {Eigenvalues and the max-cut problem},
  journal = {Czechoslovak Mathematical Journal},
  volume = {40},
  number = {2},
  pages = {343--352},
  year = {1990},
  doi = {10.21136/CMJ.1990.102386}
}

@article{EggerMarecekWoerner2021,
  author = {Egger, Daniel J. and Mare{\v{c}}ek, Jakub and Woerner, Stefan},
  title = {Warm-starting quantum optimization},
  journal = {Quantum},
  volume = {5},
  pages = {479},
  year = {2021},
  doi = {10.22331/q-2021-06-17-479}
}

@article{Tate2023WarmStart,
  author = {Tate, Reuben and Farhadi, Majid and Herold, Creston and Mohler, Greg and Gupta, Swati},
  title = {Bridging classical and quantum with {SDP}-initialized warm-starts for {QAOA}},
  journal = {ACM Transactions on Quantum Computing},
  volume = {4},
  number = {2},
  pages = {9:1--9:39},
  year = {2023},
  doi = {10.1145/3549554}
}

@article{DupontSundar2024,
  author = {Dupont, Maxime and Sundar, Bhuvanesh},
  title = {Extending relax-and-round combinatorial optimization solvers with quantum correlations},
  journal = {Physical Review A},
  volume = {109},
  number = {1},
  pages = {012429},
  year = {2024},
  doi = {10.1103/PhysRevA.109.012429}
}

@article{RendlRinaldiWiegele2010,
  author = {Rendl, Franz and Rinaldi, Giovanni and Wiegele, Angelika},
  title = {Solving {Max-Cut} to optimality by intersecting semidefinite and polyhedral relaxations},
  journal = {Mathematical Programming},
  volume = {121},
  number = {2},
  pages = {307--335},
  year = {2010},
  doi = {10.1007/s10107-008-0235-8}
}

@article{KrislockMalickRoupin2017,
  author = {Krislock, Nathan and Malick, J{\'e}r{\^o}me and Roupin, Fr{\'e}d{\'e}ric},
  title = {{BiqCrunch}: A semidefinite branch-and-bound method for solving binary quadratic problems},
  journal = {ACM Transactions on Mathematical Software},
  volume = {43},
  number = {4},
  pages = {32:1--32:23},
  year = {2017},
  doi = {10.1145/3005345}
}

@article{Spall1992,
  author = {Spall, James C.},
  title = {Multivariate stochastic approximation using a simultaneous perturbation gradient approximation},
  journal = {IEEE Transactions on Automatic Control},
  volume = {37},
  number = {3},
  pages = {332--341},
  year = {1992},
  doi = {10.1109/9.119632}
}

@inproceedings{TranMars2016,
  author = {Tran, Tony T. and Wang, Zhihui and Do, Minh and Rieffel, Eleanor G. and Frank, Jeremy and O'Gorman, Bryan and Venturelli, Davide and Beck, J. Christopher},
  title = {Explorations of quantum-classical approaches to scheduling a {Mars} lander activity problem},
  booktitle = {AAAI-16 Workshop on Planning for Hybrid Systems},
  series = {Technical Report WS-16-12},
  publisher = {AAAI Press},
  year = {2016},
  url = {https://ntrs.nasa.gov/citations/20190030474}
}

@misc{HanerBooth2024,
  author = {H{\"a}ner, Thomas and Booth, Kyle E. C. and Borujeni, Sima E. and Zhu, Elton Yechao},
  title = {Solving {QUBO}s with a quantum-amenable branch and bound method},
  year = {2024},
  eprint = {2407.20185},
  archivePrefix = {arXiv},
  primaryClass = {quant-ph}
}

@article{Hadfield2021Hamiltonians,
  author = {Hadfield, Stuart},
  title = {On the representation of {Boolean} and real functions as {Hamiltonians} for quantum computing},
  journal = {ACM Transactions on Quantum Computing},
  volume = {2},
  number = {4},
  pages = {18:1--18:21},
  year = {2021},
  doi = {10.1145/3478519}
}

@inproceedings{Huang2023QRBnB,
  author = {Huang, Wei-Hao and Matsuyama, Hiromichi and Nishimura, Kohji and Yamashiro, Yu},
  title = {Quantum Relaxation Based Branch-and-Bound Algorithm},
  booktitle = {2023 {IEEE} International Conference on Quantum Computing and Engineering ({QCE})},
  volume = {2},
  pages = {282--283},
  publisher = {IEEE},
  year = {2023},
  doi = {10.1109/QCE57702.2023.10243}
}

@inproceedings{Matsuyama2024QRBnB,
  author = {Matsuyama, Hiromichi and Huang, Wei-Hao and Nishimura, Kohji and Yamashiro, Yu},
  title = {Efficient Internal Strategies in Quantum Relaxation based Branch-and-Bound},
  booktitle = {2024 {IEEE} International Conference on Quantum Computing and Engineering ({QCE})},
  volume = {1},
  pages = {470--480},
  publisher = {IEEE},
  year = {2024},
  doi = {10.1109/QCE60285.2024.00062},
  eprint = {2405.00935},
  archivePrefix = {arXiv}
}

@misc{Fischer2024Shrinking,
  author = {Fischer, Victor and Passek, Maximilian and Wagner, Friedrich and Fin\v{z}gar, Jernej Rudi and Palackal, Lilly and Mendl, Christian B.},
  title = {Quantum and classical correlations in shrinking algorithms for optimization},
  year = {2024},
  eprint = {2404.17242},
  archivePrefix = {arXiv},
  primaryClass = {quant-ph},
  doi = {10.48550/arXiv.2404.17242}
}

@misc{Czegel2025QCBB,
  author = {Cz{\'e}gel, Andr{\'a}s and Sipos, D{\'a}vid and G.-T{\'o}th, Bogl{\'a}rka},
  title = {A Quantum-Classical Hybrid Branch \& Bound Algorithm},
  year = {2025},
  eprint = {2511.19501},
  archivePrefix = {arXiv},
  primaryClass = {quant-ph}
}

@article{WurtzLove2021,
  author = {Wurtz, Jonathan and Love, Peter},
  title = {{MaxCut} quantum approximate optimization algorithm performance guarantees for $p>1$},
  journal = {Physical Review A},
  volume = {103},
  pages = {042612},
  year = {2021},
  doi = {10.1103/PhysRevA.103.042612}
}

@website{HadfieldAPS2023,
  author = {Hadfield, Stuart and Alam, Sohaib and Brady, Lucas T. and Wang, Zhihui},
  title = {Iterative approaches to quantum optimization via problem reductions ({APS} {March Meeting} 2023, {Abstract B64.00001})},
  year = {2023},
  url = {https://meetings.aps.org/Meeting/MAR23/Session/B64.1}
}

@misc{WyboGreedy2026,
  author = {Wybo, Elisabeth and R\"onkk\"o, Jami and Hirviniemi, Olli and Fin\v{z}gar, Jernej Rudi and Leib, Martin},
  title = {A scalable quantum-enhanced greedy algorithm for maximum independent set problems},
  year = {2026},
  eprint = {2601.21923},
  archivePrefix = {arXiv},
  primaryClass = {quant-ph}
}

@misc{WyboFinzgar2026QISS,
  author = {Wybo, Elisabeth and Fin\v{z}gar, Jernej Rudi},
  title = {Quantum-informed surrogate sampling for combinatorial optimization},
  year = {2026},
  eprint = {2607.22372},
  archivePrefix = {arXiv},
  primaryClass = {quant-ph}
}

@article{Schuetz2025qReduMIS,
  author = {Schuetz, Martin J. A. and Yalovetzky, Romina and Andrist, Ruben S. and Salton, Grant and Sun, Yue and Raymond, Rudy and Chakrabarti, Shouvanik and Acharya, Atithi and Shaydulin, Ruslan and Pistoia, Marco and Katzgraber, Helmut G.},
  title = {Quantum-informed reduction algorithm for the maximum independent set problem},
  journal = {Physical Review Research},
  year = {2026},
  volume = {8},
  pages = {033296},
  eprint = {2503.12551},
  archivePrefix = {arXiv},
  doi = {10.1103/nh3b-1wv5}
}

@misc{Yalovetzky2026Portfolio,
  author = {Yalovetzky, Romina and Schuetz, Martin J. A. and He, Zichang and Shen, Jiayu and Sun, Yue and Raymond, Rudy and Sahay, Shauna and Perla, Kishore and Andrist, Ruben S. and Salton, Grant and Katzgraber, Helmut G. and Bongiovanni, Roger and Kumar, Niraj and Otter, Rob},
  title = {Quantum-informed portfolio selection: An end-to-end pipeline validated on trapped-ion hardware with real market data},
  year = {2026},
  eprint = {2607.01037},
  archivePrefix = {arXiv},
  primaryClass = {quant-ph}
}

@misc{Wang2025ExpectationComplexity,
  author = {Wang, Jingheng and Chen, Shengminjie and Sun, Xiaoming and Zhang, Jialin},
  title = {A unified complexity-algorithm account of constant-round {QAOA} expectation computation},
  year = {2025},
  eprint = {2511.20212},
  archivePrefix = {arXiv},
  primaryClass = {quant-ph}
}

@misc{Hadfield2026ExpectationComplexity,
  author = {Hadfield, Stuart},
  title = {Evaluating {QAOA} expectation values can be as hard as counting optimal solutions},
  year = {2026},
  eprint = {2608.11385},
  archivePrefix = {arXiv},
  primaryClass = {quant-ph}
}

@misc{Vercellino2025QCBP,
  author = {Vercellino, Chiara and Naghmouchi, M. Yassine and Coelho, Wesley and Vitali, Giacomo and Scionti, Alberto and Viviani, Paolo and Terzo, Olivier and Montrucchio, Bartolomeo},
  title = {Hybrid Quantum-Classical Branch-and-Price Method for the Vertex Coloring Problem},
  year = {2025},
  eprint = {2508.18887},
  archivePrefix = {arXiv},
  primaryClass = {quant-ph}
}

@misc{bassa2026iterative,
  author = {Pariente Bassa, Ruben and Quinton, Finley A. and Fuchs, Franz G. and Halffmann, Pascal},
  title = {Iterative quantum algorithms for the minimum vertex cover problem based on continuous-time quantum walks},
  year = {2026},
  eprint = {2607.27915},
  archivePrefix = {arXiv},
  primaryClass = {quant-ph},
  doi = {10.48550/arXiv.2607.27915},
  url = {https://arxiv.org/abs/2607.27915}
}

@article{lucas2014ising,
 author = {Lucas, Andrew},
 title = {Ising formulations of many {NP} problems},
 journal = {Frontiers in Physics},
 volume = {2},
 pages = {5},
 year = {2014},
 eprint = {1302.5843},
 archivePrefix = {arXiv},
 doi = {10.3389/fphy.2014.00005}
}

@article{HadfieldHoggRieffel2023,
  author = {Hadfield, Stuart and Hogg, Tad and Rieffel, Eleanor G.},
  title = {Analytical framework for quantum alternating operator ans{\"a}tze},
  journal = {Quantum Science and Technology},
  volume = {8},
  number = {1},
  pages = {015017},
  year = {2023},
  doi = {10.1088/2058-9565/aca3ce}
}
\appendix

\section{Weighted MaxCut under pairwise reductions}
\label{app:maxcut}
For completeness we record the contraction identities used in the
main text. Let $a$ be the additive constant accumulated by previous
contractions, with $a=0$ at the root. The reduced MaxCut Hamiltonian
on the remaining graph is
\begin{equation}
C(G;a)=aI+\sum_{(ij)\in E}\frac{w_{ij}}{2}(I-Z_iZ_j).
\label{eq:maxcutH}
\end{equation}
Fix $Z_1=\sigma Z_2$, where $\sigma\in\{+1,-1\}$. The edge $(1,2)$
contributes the constant $w_{12}(1-\sigma)/2$. For every neighbor
$j\neq 2$ of vertex 1,
\begin{equation}
\frac{w_{1j}}{2}(I-Z_1Z_j)
=\begin{cases}
\frac{w_{1j}}{2}(I-Z_2Z_j),&\sigma=+1,\\[3pt]
w_{1j}-\frac{w_{1j}}{2}(I-Z_2Z_j),&\sigma=-1.
\end{cases}
\end{equation}
Thus the weights incident on vertex 1 are added to those incident
on vertex 2 when $\sigma=+1$, and subtracted when $\sigma=-1$.
More explicitly, taking the weight of an absent edge to be zero,
the contraction gives
\begin{align}
w'_{2j}&=w_{2j}+\sigma w_{1j},
&&j\notin\{1,2\}, \nonumber\\
a'&=a+\frac{1-\sigma}{2}
\sum_{j:(1,j)\in E}w_{1j}.
\label{eq:maxcut-contraction}
\end{align}
All other edge weights are unchanged. The result is another
real-weighted MaxCut instance on one fewer vertex.
A reduced assignment lifts to an original assignment by reversing
the substitutions. The update of $a$ ensures that the reduced
objective equals the value of that lifted assignment, establishing
Eq.~\eqref{eq:costpreserve} for this application.

\section{Precision of quantum scores}
\label{app:precision}
For a fixed reduced state, let $\mu_e=\langle Z_iZ_j\rangle$ for each
of $K$ candidate pairs $e=\{i,j\}$. The same $N$ independent
computational basis shots yield estimates $\widehat\mu_e$ for all
pairs. Hoeffding's inequality and a union bound give %
\begin{equation}
\Pr\!\left[\max_e|\widehat\mu_e-\mu_e|>\varepsilon\right]
\leq 2K\exp(-N\varepsilon^2/2).
\label{eq:iqo-score-precision}
\end{equation}
Thus $N\geq 2\varepsilon^{-2}\log(2K/\delta)$ suffices for simultaneous
error at most $\varepsilon$ with probability at least $1-\delta$.
On this event, the leading pair under $|\mu_e|$ is unambiguous when its
magnitude exceeds every competitor by more than $2\varepsilon$; its
preferred sign is stable when $|\mu_e|>\varepsilon$.
Resolving these margins is a condition for reproducing a quantum
preference, not for certifying the optimum. The controller may break
unresolved ties or use a classical score while retaining both parity
outcomes. A measured candidate contributes its classically evaluated
objective, rather than an estimated mean, to the incumbent. Consequently,
sampling error changes the order and amount of search, while valid node
bounds, lifting, and coverage retain their original correctness roles.

\section{Proofs for the greedy and backtracking separation}
\label{app:greedytrap}
We first prove Theorem~\ref{thm:greedy-rqaoa-trap}. If $u,v,w$ lie on the same side of a cut of $G_m$, placing every $b_j$ on the opposite side cuts all $3m$ bipartite edges and neither internal edge, giving value $3m$. If instead $u,v,w$ are split $2$--$1$, each $b_j$ can cut at most two of its three incident bipartite edges. The bipartite contribution is therefore at most $2m$, while the path $u-v-w$ contributes at most two. For $m\geq3$, $2m+2<3m$, so
\begin{equation}
\operatorname{OPT}(G_m)=3m,
\qquad
Z_uZ_v=Z_vZ_w=+1
\label{eq:trap-opt-app}
\end{equation}
for every optimal cut.
It remains to show that optimized level-$1$ QAOA nevertheless assigns the opposite sign to the two internal edges. We use the standard convention
\begin{equation}
|\psi(\beta,\gamma)\rangle=e^{-i\beta\sum_i X_i}e^{-i\gamma C}|+\rangle^{\otimes n}.
\end{equation}
Write
\begin{equation}
c=\cos\gamma,\qquad s=\sin\gamma,\qquad t=\cos(2\gamma).
\end{equation}
For an edge $(i,j)$ let $\delta_i=\deg(i)-1$, $\delta_j=\deg(j)-1$, and let $\lambda_{ij}=|N(i)\cap N(j)|$ be the number of triangles containing the edge. The standard level-$1$ MaxCut calculation gives~\cite{Wang2018,Ozaeta2022}
\begin{align}
\langle C_{ij}\rangle
&=
\frac12+\frac14\sin(4\beta)s
\left(c^{\delta_i}+c^{\delta_j}\right)
\nonumber\\
&\quad
-\frac14\sin^2(2\beta)
c^{\delta_i+\delta_j-2\lambda_{ij}}
\left(1-t^{\lambda_{ij}}\right).
\label{eq:p1-edge-general}
\end{align}
There are three edge types. For an internal edge $(u,v)$ or $(v,w)$,
\begin{equation}
(\delta_i,\delta_j,\lambda_{ij})=(m,m+1,m);
\end{equation}
for $(u,b_j)$ or $(w,b_j)$,
\begin{equation}
(\delta_i,\delta_j,\lambda_{ij})=(m,2,1);
\end{equation}
and for $(v,b_j)$,
\begin{equation}
(\delta_i,\delta_j,\lambda_{ij})=(m+1,2,2).
\end{equation}
Summing Eq.~\eqref{eq:p1-edge-general} over all $3m+2$ edges gives
\begin{equation}
F_m(\beta,\gamma)
=
\frac{3m+2}{2}
+A_m(\gamma)\sin(4\beta)
-B_m(\gamma)\sin^2(2\beta),
\label{eq:trap-p1-objective}
\end{equation}
where
\begin{align}
A_m(\gamma)
&=
\frac{s}{4}
\left[
2(m+1)c^m+(m+2)c^{m+1}+3mc^2
\right],
\label{eq:trap-Am}\\
B_m(\gamma)
&=\frac14\Big[
2c(1-t^m)+2mc^m(1-t)
\nonumber\\
&\qquad\quad+mc^{m-1}(1-t^2)\Big].
\label{eq:trap-Bm}
\end{align}
The optimization over $\beta$ is exact:
\begin{equation}
\max_\beta\left\{
A\sin(4\beta)-B\sin^2(2\beta)
\right\}
=
-\frac{B}{2}+\sqrt{A^2+\frac{B^2}{4}}.
\label{eq:trap-beta-opt}
\end{equation}
We now characterize the global maximizers as $m\to\infty$. In the asymptotic equations below, an arrow annotated by $m\to\infty$ denotes convergence along a sequence of globally maximizing QAOA parameters. It is unrelated to the reduction arrows in Eq.~\eqref{eq:path}. Values with $|c|\to1$ cannot maximize Eq.~\eqref{eq:trap-p1-objective}: there $|s|\to0$ and the gain over the random-cut value is $o(m)$, whereas fixed $c^2\in(0,1)$ gives a gain of order $m$. Hence every maximizing sequence has an accumulation point away from $|c|=1$. On compact subsets away from these endpoints, the $c^m$ terms vanish uniformly. The term $c\,t^m$ also vanishes uniformly, with its prefactor controlling the neighborhood of $c=0$, so
\begin{align}
\frac{A_m}{m}&=\frac34sc^2+o(1),\\
B_m&=\frac{c}{2}+o(1).
\label{eq:trap-asymptotics-AB}
\end{align}
After Eq.~\eqref{eq:trap-beta-opt}, the leading optimization is therefore to maximize
\begin{equation}
|s|c^2=\sqrt{1-c^2}\,c^2,
\end{equation}
whose only maximizing values satisfy
\begin{equation}
c^2\xrightarrow[m\to\infty]{}\frac23.
\label{eq:trap-csquare}
\end{equation}
The sign of the limiting branch requires a uniform comparison, because the maximizing value of $c$ may itself depend on $m$. Define
\begin{equation}
\Phi_m(c):=-\frac{B_m(c)}{2}+\sqrt{A_m(c)^2+\frac{B_m(c)^2}{4}},
\end{equation}
where the sign of $s$ is immaterial after optimizing over $\beta$.
\begin{lemma}[Uniform branch selection]
\label{lem:uniform-branch}
Let $c_\pm=\pm\sqrt{2/3}$. There are disjoint fixed neighborhoods $N_\pm$ of $c_\pm$ such that every global maximizer of $\Phi_m$ lies in $N_+\cup N_-$ for all sufficiently large $m$. Uniformly on these neighborhoods,
\begin{equation}
\Phi_m(c)=\frac{3m}{4}c^2\sqrt{1-c^2}-\frac{c}{4}
+O(m^{-1})+O(mq^m)
\label{eq:uniform-Phi}
\end{equation}
for some $q<1$. The maximizer within each $N_\pm$ is $c_\pm+O(m^{-1})$, and the optimized value in $N_-$ exceeds that in $N_+$ by $\frac12\sqrt{2/3}+o(1)$.
\end{lemma}
\begin{proof}
On compact subsets of $(-1,1)$ containing $N_+\cup N_-$, the exponentially small terms in Eqs.~\eqref{eq:trap-Am}--\eqref{eq:trap-Bm} and their first two derivatives are uniform. Moreover $|A_m|=\Theta(m)$ on these neighborhoods, so expanding the square root in Eq.~\eqref{eq:trap-beta-opt} gives Eq.~\eqref{eq:uniform-Phi}. The leading function $g(c)=c^2\sqrt{1-c^2}$ has exactly two global maxima, at $c_\pm$, and $g''(c_\pm)=-4\sqrt3<0$. On the fixed compact complement of $N_+\cup N_-$, the bound $\Phi_m\le |A_m|+|B_m|$ gives an $\Omega(m)$ deficit relative to these neighborhoods, proving localization of every global maximizer. The stationary equation derived from Eq.~\eqref{eq:uniform-Phi} moves each local maximizer by only $O(m^{-1})$. Evaluating the $O(1)$ term $-c/4$ at $c_-$ and $c_+$ gives the stated gap; the displacement changes the value only by $O(m^{-1})$.
\end{proof}
Lemma~\ref{lem:uniform-branch} removes the possible competition between $m$-dependent perturbations of the two leading maximizers. Consequently every sequence of global maximizers satisfies
\begin{equation}
\cos\gamma_\star\xrightarrow[m\to\infty]{}-\sqrt{\frac23}.
\label{eq:trap-gamma-limit}
\end{equation}
The optimizing condition in Eq.~\eqref{eq:trap-beta-opt} also yields
\begin{equation}
\sin(4\beta_\star)\sin\gamma_\star\xrightarrow[m\to\infty]{}\frac{1}{\sqrt3},
\qquad
\sin^2(2\beta_\star)\xrightarrow[m\to\infty]{}\frac12.
\label{eq:trap-beta-limits}
\end{equation}
Using $\langle Z_iZ_j\rangle=1-2\langle C_{ij}\rangle$ in Eq.~\eqref{eq:p1-edge-general}, Eqs.~\eqref{eq:trap-gamma-limit}--\eqref{eq:trap-beta-limits} give for either internal edge
\begin{equation}
\rho_{\rm int}
:=
\langle Z_uZ_v\rangle
=
\langle Z_vZ_w\rangle
\xrightarrow[m\to\infty]{}
-\frac14\sqrt{\frac23},
\label{eq:trap-rho-internal}
\end{equation}
whereas both bipartite edge types satisfy
\begin{equation}
\rho_{\rm bip}\xrightarrow[m\to\infty]{}-\frac{1}{3\sqrt3}.
\label{eq:trap-rho-bip}
\end{equation}
Since
\begin{equation}
\frac14\sqrt{\frac23}>\frac{1}{3\sqrt3},
\end{equation}
the two internal edges have strictly larger correlation magnitude than every bipartite edge for all sufficiently large $m$, and their correlations are negative. If infinitely many $m$ had a global optimizer violating this conclusion, a subsequence would contradict the limits above. This proves the existence of $m_0$ in Theorem~\ref{thm:greedy-rqaoa-trap}.
A numerical scan of the closed-form expression through $m=500$ is consistent with the asymptotic result and finds a contiguous failing tail from $m=12$ onward. Near the start of that tail the competing optimizer branches can be close, so a practical finite-precision optimizer may land on the benign branch; the theorem concerns global optimization. %
The complete finite-size scan was retained with the numerical analysis.
By symmetry suppose the first reduction imposes $Z_uZ_v=-1$. If $w$ lies on the same side as $u$, the vertices $u,v,w$ form a $2$--$1$ split, each $b_j$ contributes at most two cut edges, and both internal edges can be cut. Hence
\begin{equation}
\operatorname{OPT}(G_m\mid Z_uZ_v=-1)=2m+2.
\label{eq:trap-wrong-branch}
\end{equation}
If instead $w$ lies on the same side as $v$, the value is only $2m+1$. Thus no later sequence of reductions can repair the greedy branch and Eq.~\eqref{eq:trap-ratio-main} follows.
The complementary child $Z_uZ_v=+1$ contains the global optimum. After contracting $u$ and $v$, only the parity of the contracted vertex relative to $w$ remains relevant. Choosing them equal and placing every $b_j$ opposite cuts all $3m$ bipartite edges. Choosing them opposite gives at most $2m+1$. The residual choice separates over the $m$ vertices $b_j$, so it can be solved in $O(m)$ time. One root backtrack therefore suffices to recover the exact optimum.
The construction above is specific to $p=1$ and the standard all-edge RQAOA selection rule. A complementary locality obstruction shows that moving to another fixed QAOA depth cannot in general make a locally inferred parity safe to fix irreversibly. For an edge $e=(i,j)$ write $Z_e:=Z_iZ_j$, and let $\rho_e^{(p)}$ denote the two-qubit reduced state of the standard depth-$p$ QAOA output on its endpoints.
\begin{proposition}[Fixed-depth local-state ambiguity]
\label{prop:fixed-depth-ambiguity}
For every fixed QAOA depth $p\geq1$ and every $W>1$, there is a connected weighted MaxCut instance $H_{p,W}$ containing edges $e_-$ and $e_+$ whose rooted weighted depth-$p$ neighborhoods are isomorphic, with induced endpoint relabeling $P_\phi$, such that
\begin{equation}
\rho_{e_-}^{(p)}=P_\phi\,\rho_{e_+}^{(p)}\,P_\phi^\dagger
\label{eq:local-identical-state}
\end{equation}
for every choice of the standard graph-global QAOA layer angles. Thus all corresponding edge-local measurement statistics, including $\langle Z_{e_-}\rangle=\langle Z_{e_+}\rangle$, agree, while every globally optimal cut requires
\begin{equation}
Z_{e_-}=-1,\qquad Z_{e_+}=+1.
\label{eq:local-opposite-opt}
\end{equation}
Consequently no mandatory isomorphism-equivariant parity rule whose input is only the complete edge-local QAOA state can be correct on both edges. More quantitatively, if one of $e_-$ and $e_+$ is selected uniformly at random, any such rule---deterministic or randomized---has mean parity-success probability exactly $1/2$, because its output distribution is identical on the two locally indistinguishable inputs while the required parities are opposite. In particular, if the IQO candidate feature set is $\{Z_{e_-},Z_{e_+}\}$, randomly breaking their correlation-magnitude tie and fixing the common preferred sign makes an incorrect first reduction with probability $\frac{1}{2}$ (with any fixed common sign convention if the correlator vanishes).
Let $\mathrm{OPT}_{\mathrm{wrong}}(H_{p,W})$ be the larger
of the two optimal cut values obtained by fixing
$Z_{e_-}=+1$ or $Z_{e_+}=-1$. An incorrectly selected
branch can attain no more than this value. As shown below,
\begin{equation}
\begin{aligned}
\frac{\mathrm{OPT}_{\mathrm{wrong}}(H_{p,W})}
{\mathrm{OPT}(H_{p,W})}
&=\frac{(4p+4)W+3}{(4p+5)W+2}\\
&\xrightarrow[W\to\infty]{}\frac{4p+4}{4p+5}.
\end{aligned}\label{eq:fixed-p-ratio}
\end{equation}
One backtrack to the complementary parity choice restores a branch containing a global optimum.
\end{proposition}
\begin{proof}
Take an even cycle $Q_-=C_{2p+4}$ and an odd cycle $Q_+=C_{2p+3}$. In each cycle choose a distinguished edge, $e_-$ and $e_+$, of weight $1$, and give every other cycle edge weight $W>1$. Connect the two cycles by a weight-$1$ bridge whose endpoints lie at distance $p+1$ from the corresponding distinguished edge. The resulting graph $H_{p,W}$ is connected.
For any two-qubit operator $O$ on a distinguished edge, its depth-$p$ Heisenberg causal cone is contained in the weighted depth-$p$ neighborhood: mixer layers do not enlarge support and each commuting MaxCut phase layer enlarges it by at most one graph step~\cite{FarhiGamarnikGutmann2020}. The rooted neighborhoods of $e_-$ and $e_+$ are weight-preserving isomorphic, each a central weight-$1$ edge with two length-$p$ weight-$W$ arms; the cycle closures and bridge lie outside the cone. The gates appearing in the two causal cones are therefore related by the induced qubit permutation, while $|+\rangle^{\otimes n}$ is permutation invariant. Corresponding expectations agree for every $O$, so the two-qubit Pauli basis gives Eq.~\eqref{eq:local-identical-state}. The identity is pointwise in the graph-global layer angles and hence survives arbitrary classical selection of those angles using the full instance.
The optimal parities are nevertheless opposite. The even cycle is bipartite, so every optimum cuts $e_-$ and all $2p+3$ weight-$W$ edges,
\begin{equation}
\operatorname{OPT}(Q_-)=(2p+3)W+1.
\end{equation}
For the odd cycle, at least one edge must remain uncut. Since $e_+$ is the unique minimum-weight edge, every optimum leaves precisely $e_+$ uncut and cuts all $2p+2$ weight-$W$ edges,
\begin{equation}
\operatorname{OPT}(Q_+)=(2p+2)W.
\end{equation}
A global spin flip of either cycle preserves its internal cut, so their relative orientations can always be chosen to cut the bridge. Thus
\begin{equation}
\operatorname{OPT}(H_{p,W})=(4p+5)W+2.
\end{equation}
The general local-rule consequence now follows directly from Eq.~\eqref{eq:local-identical-state}: isomorphism equivariance forces the same output distribution on the two identified local states, whereas Eq.~\eqref{eq:local-opposite-opt} makes the two correct parity labels opposite. Averaging over a uniformly selected distinguished edge therefore gives success probability $1/2$ for any mandatory parity rule based only on that local state. Because the two correlators are identical, the correlation-based special case assigns the same preferred sign to both. Exactly one of the two optimal relations in Eq.~\eqref{eq:local-opposite-opt} agrees with that sign, so random tie breaking between the two edges makes a wrong reduction with probability $\frac{1}{2}$. If $e_-$ is incorrectly fixed to $+1$, its even cycle can cut at most $2p+2$ weight-$W$ edges, and
\begin{equation}
\operatorname{OPT}(H_{p,W}\mid Z_{e_-}=+1)=(4p+4)W+1.
\end{equation}
If $e_+$ is incorrectly fixed to $-1$, its odd cycle can cut at most $2p+1$ weight-$W$ edges in addition to $e_+$, giving
\begin{equation}
\operatorname{OPT}(H_{p,W}\mid Z_{e_+}=-1)=(4p+4)W+3.
\end{equation}
Equation~\eqref{eq:fixed-p-ratio} follows. Reversing the selected parity restores a branch containing a global optimum, and the residual graph is a pair of weighted cycles joined by a bridge, hence has constant treewidth and can be solved exactly in linear time.
\end{proof}
Thus complete two-qubit tomography returns identical local QAOA evidence for the two relations under the rooted identification: no mandatory isomorphism-equivariant parity rule using only this edge-local state is sound on both, although abstaining or branching remains safe. Explicitly edge-dependent parameter choices or multi-angle ansatzes are outside the statement. For $p=2$ the construction has $15$ vertices and $\operatorname{OPT}_{\rm wrong}/\operatorname{OPT}\to12/13$. Unlike Theorem~\ref{thm:greedy-rqaoa-trap}, which gives a strict all-edge ranking, Proposition~\ref{prop:fixed-depth-ambiguity} restricts the candidate set and uses the forced tie; both support treating fixed-depth quantum information as branch guidance rather than a logical certificate.
\section{Numerical implementation}
\label{app:numerics}
\paragraph{Tree-IQO simulations.}
The Tree-IQO simulations use exact state vectors at depth $p=1$. At each reduced problem we evaluate a $7\times7$ grid over $\gamma\in[0,\pi)$ and $\beta\in[0,\pi/2)$ and refine the best point with Nelder--Mead, using at most 80 iterations, parameter tolerance $5\times10^{-4}$, and objective tolerance $5\times10^{-5}$. These angle intervals are simply the optimizer convention used for all signed reduced instances. We do not assume they are a symmetry-reduced fundamental domain. The angle search itself uses exact expectation values. Reported shot counts refer only to the subsequent finite-shot sampling and are not end-to-end hardware training budgets. Throughout the paper, ``noiseless correlators'' means exact state-vector expectation values at the numerically optimized angles, not a claim that those angles are globally optimal. In Fig.~\ref{fig:treeprune}(a), eight computational basis samples are drawn from each optimized state, used to estimate all edge correlations, and also lifted and evaluated as candidate cuts.
For Fig.~\ref{fig:treeprune}(b), the QAOA correlators are evaluated exactly from the simulated state so that only the ordering is varied. The same IQO pass draws eight samples per optimized state. Those strings contribute to a common incumbent but not to the branch scores. The classical DEC baseline~\cite{HassinLeshenko2021} chooses the present pair of largest edge weight, prefers $\sigma=-1$, and uses the contraction direction with larger total reduced edge weight. Both parities are still explored in the certifying tree, so DEC supplies only the pair order, preferred child, and orientation convention.
The random baseline chooses the pair uniformly and uses the same difference-contraction convention. All three searches start from the same incumbent, the better of the noiseless-correlator IQO leaf and its retained samples. The random curve averages three independently generated schedules per instance.
The stored schedule is then replayed classically with both parities at every branch. We evaluate Eq.~\eqref{eq:maxcutub} before expanding descendants and count a state as visited when its bound is evaluated, including states pruned immediately. Terminal states are counted before their residual problem is solved exactly. The code checks cost preservation under signed contractions, and all finite-shot runs at the reported sizes were verified against brute-force MaxCut. Numerical settings are exposed as command-line arguments rather than hard-coded into the search logic.
The state-vector tests use double precision, discard contracted edge weights of magnitude at most $10^{-12}$, and prune when the computed bound is at most $c_{\rm best}+10^{-12}$; brute-force checks use absolute tolerance $10^{-8}$. Spectral eigenvalues are computed without interval enclosures. Numerical certification here and below means completion of the floating-point search with the stated cross-checks. Rigorous certificates additionally require conservative enclosures of objective and bound errors, including discarded coefficients and a lower enclosure of $\lambda_{\min}(A)$ in Eq.~\eqref{eq:spectralub}.
Table~\ref{tab:numconfigs} summarizes the matching conventions for the numerical panels. The main and extension studies use the same complete-weighted ensemble but independent deterministic instance streams, so numbers from different rows need not coincide. For example, the $9.4\%$ value at $n=14$ in Fig.~\ref{fig:treeprune}(b) and the $9.5\%$ exact-correlator value in Appendix~\ref{app:bound-shot} come from different streams and use different common incumbents.
\begin{table*}[t]
\caption{Configuration map for the Tree-IQO numerical panels. ``Main'' and ``extension'' are independent deterministic draws from the same complete weighted ensemble, $w_{ij}\stackrel{\rm iid}{\sim}U[0.2,1]$. Each row is internally matched. Cross-row equality is not implied. ``Leaf+samples'' denotes the better of the indicated IQO leaf and retained candidate samples.}
\label{tab:numconfigs}
\centering
\small
\begin{tabular}{@{}p{0.12\textwidth}p{0.10\textwidth}p{0.22\textwidth}p{0.31\textwidth}p{0.15\textwidth}@{}}
\toprule
Result & instance stream & reduction schedule & initial incumbent & node bound \\
\midrule
Fig.~\ref{fig:treeprune}(a) & main & 8-shot IQO & leaf vs. leaf+samples & elementary \\
Fig.~\ref{fig:treeprune}(b) & main & noiseless corr. IQO / DEC / random & common noiseless corr. leaf+samples & elementary \\
Bound test (App.~\ref{app:bound-shot}) & extension & 8-shot IQO & leaf+samples & elem./spectral \\
Shot-count test (App.~\ref{app:bound-shot}) & extension & 8/32/128/noiseless IQO & common noiseless corr. IQO leaf & elementary \\
Table~\ref{tab:hybridcontrol} & extension & 8-shot IQO / DEC & method-specific; hybrid adds IQO samples & spectral \\
Table~\ref{tab:training-reuse} & SPSA & 16-it finite-shot SPSA IQO & final vs. training+final samples & elementary \\
Table~\ref{tab:sparse} & sparse & 8-shot IQO & leaf vs. leaf+samples & elem./spectral \\
Sparse ordering (App.~\ref{app:sparse}) & sparse & noiseless corr. IQO / DEC / random & common noiseless corr. leaf+samples & elementary \\
Table~\ref{tab:analytic-ordering} & analytic-$p=1$ & noiseless corr. IQO / DEC / random & common noiseless corr. IQO leaf & elementary \\
\bottomrule
\end{tabular}
\end{table*}
For the hybrid control in Appendix~\ref{app:classicalbaselines}, we keep the classical DEC schedule and spectral bound fixed and change only the initial incumbent. The quantum-assisted version replaces the DEC leaf incumbent by the better of that leaf and the best bit string among the eight-shot $p=1$ IQO sample batches generated on the matched instance. Thus any node-count change in this control comes solely from adding the already generated quantum candidate pool to an otherwise classical search.
\paragraph{Statistical reporting.}
The statistical unit is the problem instance. For finite-shot results, one instance-level observation includes the randomness of the sampled run on that instance, so the error bars describe variation across problem instances and their associated draws. Figure~\ref{fig:treeprune} shows one standard error of the across-instance mean, $s/\sqrt{50}$; the bound and shot-count tests in Appendix~\ref{app:bound-shot} use the same convention. Table~\ref{tab:sparse} uses $s/\sqrt{30}$. The SPSA control in Table~\ref{tab:training-reuse} uses 30 matched instances per size and the same $50{,}000$-resample paired bootstrap for differences quoted in the text. For the random-order curve in Fig.~\ref{fig:treeprune}(b), the three random schedules are averaged within each instance before the across-instance standard error is computed. Other direct matched comparisons use 95\% percentile-bootstrap intervals from $50{,}000$ resamples of the per-instance differences. We use these intervals as uncertainty summaries, without multiplicity correction.
Fixed pseudorandom seeds and raw instance-level data were retained for the reported runs.
\section{Additional Tree-IQO tests}
\label{app:extra_numerics}
The tests below use independent instance streams from the main benchmark,
as specified in Table~\ref{tab:numconfigs}. For the $n\leq14$ ablations,
brute force verifies the returned optima but does not guide the searches.
Matched uncertainty intervals use the bootstrap procedure of
Appendix~\ref{app:numerics}.
\subsection{Training-sample reuse and classical controls}
\label{app:training-reuse}
\begin{figure*}[t]
\centering
\includegraphics[width=0.93\textwidth]{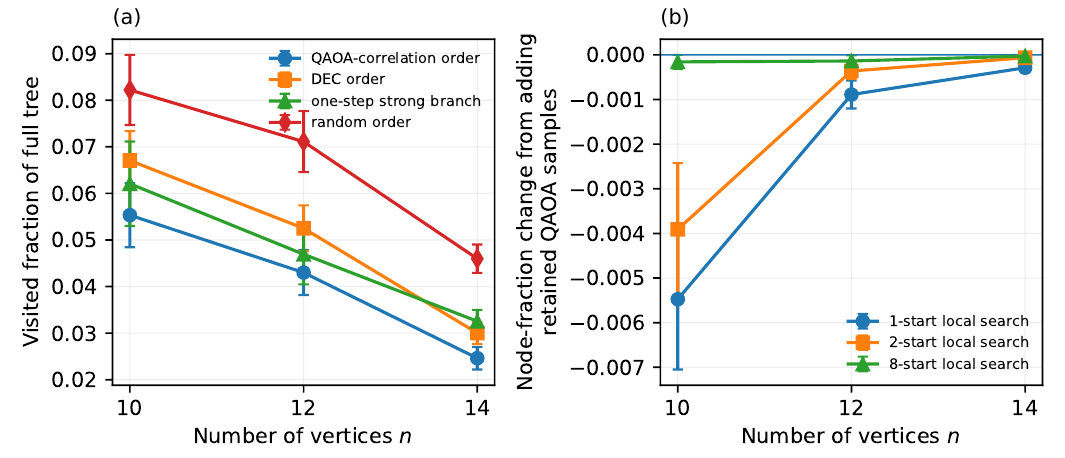}
\caption{Matched classical controls. (a) Visited fraction with a
common IQO-leaf incumbent and spectral bound, 20 complete weighted
instances per size. The strong brancher's schedule-construction
cost is excluded from the vertical axis. (b) Change in visited
fraction when retained QAOA samples augment one-, two-, or
eight-start one-flip local search before fixed DEC search,
50 instances per size. Negative values mean fewer nodes;
error bars are one standard error.}
\label{fig:classical-controls}
\end{figure*}

The main benchmark optimizes QAOA angles from exact expectations and
reuses only the final correlation-estimation shots. To test training-shot
reuse, we also train every IQO node with 16 iterations of simultaneous
perturbation stochastic approximation (SPSA)~\cite{Spall1992}. We fix
$\theta_0=(\gamma_0,\beta_0)=(0.55,\pi/8)$ for every instance and signed
representation. At iteration $k=0,\ldots,15$ the gains are
\begin{equation}
a_k=\frac{0.20}{(k+5)^{0.602}},\qquad
c_k=\frac{0.12}{(k+1)^{0.101}}.
\end{equation}
Two evaluations at $\theta_k\pm c_k\Delta_k$, with independent
Rademacher $\Delta_k\in\{\pm1\}^2$, use 16 shots each to estimate the
two-sided SPSA gradient; we update in the ascent direction with gain
$a_k$. Angles are wrapped to $\gamma\in[0,\pi)$ and
$\beta\in[0,\pi/2)$ after each update. This is an optimizer convention
for the signed instances, not a claimed symmetry of their landscapes.
There is no early stopping or accept/reject step. A final 32-shot batch
at the learned angles determines the next reduction.
On 30 independent complete weighted instances per
$n\in\{10,12,14\}$, we replay each learned schedule with either the
final batches alone or both final and training shots as incumbent
candidates. Each retained string is lifted and evaluated on the
original instance. Holding the schedule fixed isolates candidate
reuse (Table~\ref{tab:training-reuse}). The runs generate
$4608,5632,6656$ training strings and $288,352,416$ final-batch
strings, respectively. The training shots are part of the optimizer
cost. Reusing them requires no additional measurements.
\begin{table*}[t]
\caption{Training-sample reuse on 30 matched complete weighted MaxCut
instances per size. Final retains only the final correlation batches;
adding training also retains SPSA shots. AR is the approximation ratio.
Pool/space is the raw number of all shots divided by the
$2^{n-1}$ distinct cuts up to global spin flip, not a unique-coverage
fraction. Opt. hit is the fraction of instances whose training
shots contain an optimum. Other entries are across-instance means.}
\label{tab:training-reuse}
\centering
\small
\begin{tabular}{c@{\quad}cc@{\quad}cc@{\quad}cc}
\toprule
$n$ & final AR & +training AR & final nodes & +training nodes & pool/space & opt. hit \\
\midrule
10 & 0.98168 & 0.99968 & 209.3 & 189.1 & 9.56 & 97\% \\
12 & 0.97877 & 0.99724 & 628.3 & 571.4 & 2.92 & 73\% \\
14 & 0.97340 & 0.99377 & 2185.5 & 2033.5 & 0.86 & 37\% \\
\bottomrule
\end{tabular}
\end{table*}
For $n=10,12,14$, paired 95\% bootstrap intervals for the AR gain
are $[0.0123,0.0242]$, $[0.0129,0.0242]$, and
$[0.0159,0.0250]$; intervals for the change in visited-tree
fraction, in percentage points, are $[-2.74,-1.29]$,
$[-2.03,-0.83]$, and $[-1.25,-0.64]$. At $n=10,12$ the raw sample
pool exceeds the cut space, so the near-unity full-pool incumbent
should not be extrapolated. Replaying with random 1\% and 10\%
subsets of the training shots, while keeping the final batches and
schedules fixed, also improves incumbents and reduces node counts
relative to using final batches alone.
\paragraph{Classical incumbents.}
On 50 matched extension-stream instances per size, one steepest
one-flip local-search start uses on average $33.0,45.6,63.4$
objective evaluations at $n=10,12,14$. Its mean ARs are
$0.973,0.972,0.974$, compared with $0.979,0.980,0.977$
for the matched noiseless-correlator IQO leaf plus eight-shot
samples; paired 95\% intervals for the differences include zero.
Two local-search starts reach $0.987,0.991,0.987$.
Thus retained QAOA samples can help when already measured, while
simple classical search gives strong incumbents.
\paragraph{Classical branching and combined incumbents.}
On 20 complete weighted instances per size, a one-step strong
brancher probes each available pair under both parities with the
spectral bound, chooses the pair minimizing the worse-child bound,
and explores the larger-bound child first. With a common IQO-leaf
incumbent, QAOA, strong branching, DEC, and random ordering have
the node counts shown in Fig.~\ref{fig:classical-controls}(a).
The QAOA versus strong-branching comparison is unresolved at
$n=10,12$ and favors QAOA at $n=14$ in this sample.
Constructing the strong-branching schedules itself costs another
$330,572,910$ child-bound evaluations on average.
In a separate matched control, we add the eight-shot QAOA candidates
to one-, two-, or eight-start one-flip local search before a fixed DEC
search using the spectral bound. Relative to one start alone, the
combined incumbent lowers the mean visited fraction by
$0.547,0.089,0.029$ percentage points at $n=10,12,14$; paired 95\%
intervals exclude zero. The added value diminishes with stronger
classical incumbents [Fig.~\ref{fig:classical-controls}(b)].
As an angle-search check, a $9\times9$ grid followed by three tighter
independent refinements changed neither the selected root pair nor
its parity relative to the usual $7\times7$ grid and one
Nelder--Mead refinement on any of 60 roots (20 at each size).
Mean root-objective improvements stayed below $6\times10^{-7}$;
this does not establish global angle optimality.
\subsection{Bounds and finite-shot ordering}
\label{app:bound-shot}
On 50 complete weighted instances per size, we also test a stronger
node bound and the shots used to order branches. For a reduced graph
with weighted adjacency matrix $A$, signed total edge weight
$W=\sum_{i<j}w_{ij}$, and $N=|V|$ remaining vertices,
\begin{equation}
C(z)=a+\frac{W}{2}-\frac14 z^{T}Az.
\end{equation}
Since $z^TAz\ge N\lambda_{\min}(A)$ for
$z\in\{\pm1\}^{N}$, a spectral upper bound of Mohar--Poljak type is~\cite{MoharPoljak1990}
\begin{equation}
\UB_{\rm spec}(G;a)=a+\frac{W}{2}
-\frac{N}{4}\lambda_{\min}(A).
\label{eq:spectralub}
\end{equation}
Pruning with $\min\{\UB,\UB_{\rm spec}\}$ lowers the mean visited
fraction from $18.5\%$ to $9.6\%$, $14.6\%$ to $6.8\%$, and
$11.7\%$ to $5.0\%$ at $n=10,12,14$, with fewer nodes on all
150 instances. A root-level NumPy microbenchmark found the
spectral calculation about $8$--$9$ times slower per call than
the elementary bound. Fewer nodes do not imply less runtime.
To isolate ordering noise, we hold each instance's initial
noiseless-IQO leaf incumbent fixed and vary only the shots used
to estimate correlations. At $n=14$, using 8, 32, and 128 shots per
IQO step gives mean visited fractions $11.0\%$, $9.8\%$, and
$9.4\%$, versus $9.5\%$ with exact correlators. At 128 shots
the paired 95\% interval for the difference contains zero at
all three sizes. Node counts need not vary monotonically with
shot count for individual schedules.
\subsection{Analytic $p=1$ larger-$n$ ordering stress test}
\label{app:analytic-ordering}
\begin{table}[!t]
\caption{Expanded noiseless-correlator ordering test: mean visited
fraction $\pm$ one standard error. All methods share the first
IQO-leaf incumbent and elementary bound.}
\label{tab:analytic-ordering}
\centering
\scriptsize
\setlength{\tabcolsep}{3.0pt}
\begin{tabular}{crccc}
\toprule
$n$ & instances & IQO order & DEC order & random order \\
\midrule
16 & 48 & $7.10\pm0.31\%$ & $7.69\pm0.28\%$ & $9.78\pm0.26\%$ \\
18 & 48 & $5.37\pm0.22\%$ & $5.99\pm0.20\%$ & $8.25\pm0.20\%$ \\
20 & 48 & $4.29\pm0.20\%$ & $5.23\pm0.20\%$ & $7.35\pm0.19\%$ \\
22 & 48 & $3.33\pm0.17\%$ & $4.23\pm0.15\%$ & $6.37\pm0.14\%$ \\
24 & 30 & $2.45\pm0.14\%$ & $3.60\pm0.14\%$ & $5.66\pm0.16\%$ \\
26 & 12 & $1.80\pm0.19\%$ & $2.85\pm0.30\%$ & $5.16\pm0.24\%$ \\
28 & 6 & $1.55\pm0.20\%$ & $2.28\pm0.29\%$ & $4.66\pm0.20\%$ \\
\bottomrule
\end{tabular}
\end{table}
For noiseless $p=1$ ordering, we evaluate the analytic weighted QAOA
objective and edge-correlator formulas rather than simulate state
vectors. Cross-checks on random weighted complete graphs found
maximum objective and correlator discrepancies of
$1.8\times10^{-15}$ and $1.1\times10^{-15}$, respectively.
We use the same $7\times7$ grid and Nelder--Mead refinement as in
the main test.
Table~\ref{tab:analytic-ordering} gives the expanded results for
independent complete graphs with i.i.d.\ weights $U[0.2,1]$.
Each IQO pass supplies a fixed relation schedule and first leaf;
the leaf is the common incumbent for IQO, DEC, and random ordering.
No samples are used. All searches explore both parities and use
the elementary bound; three random schedules are averaged within
each instance. Every replay returned the same numerical optimum, with maximum
within-instance floating-point spread $8.6\times10^{-14}$.
The $n=26,28$ cohorts are small scale probes.

Paired 95\% intervals for IQO minus random and IQO minus DEC are
negative at every size. IQO visits fewer nodes than random on
239 of the 240 instances and fewer than DEC on 184. These intervals
are descriptive uncertainty summaries without multiplicity
correction. The decreasing visited fraction is not a runtime
comparison: schedule construction and per-node work are omitted,
the ensemble is specific, and the two largest-$n$ cohorts are small.
\subsection{Sparse-graph stress test}
\label{app:sparse}
We repeat the fixed-order experiment on 30 independent weighted
random 3-regular graphs at each $n\in\{10,12,14\}$, with independent
edge weights $U[0.2,1]$. The $p=1$ IQO pass uses eight shots per
reduction step for correlation ordering; the subsequent classical
search certifies the optimum. Table~\ref{tab:sparse} reports
means and standard errors.

\begin{table}[!t]
\caption{Sparse weighted 3-regular graphs, 30 instances per size
(mean $\pm$ one standard error). Samples uses the better of the
first IQO leaf and retained eight-shot candidates. Search columns
give the fraction of the full $2^n-1$ history tree visited.}
\label{tab:sparse}
\centering
\scriptsize
\setlength{\tabcolsep}{2.0pt}
\resizebox{\columnwidth}{!}{%
\begin{tabular}{cccccc}
\toprule
$n$ & leaf AR & sample AR & \shortstack{elem.\\leaf} & \shortstack{elem.\\samples} & \shortstack{+spec.\\samples} \\
\midrule
10 & $0.983\pm.004$ & $0.993\pm.003$ & $3.54\pm.31\%$ & $3.26\pm.30\%$ & $2.48\pm.22\%$ \\
12 & $0.975\pm.007$ & $0.996\pm.002$ & $1.69\pm.21\%$ & $1.56\pm.21\%$ & $0.98\pm.10\%$ \\
14 & $0.967\pm.005$ & $0.992\pm.003$ & $0.785\pm.090\%$ & $0.728\pm.088\%$ & $0.464\pm.050\%$ \\
\bottomrule
\end{tabular}%
}
\end{table}
The retained samples improve the mean incumbent and lower the
elementary-bound node fraction at all three sizes; paired 95\%
intervals exclude zero for both changes. The spectral bound lowers
the node fraction further. With a common incumbent and elementary
bound, noiseless-correlator IQO, DEC, and random ordering visit
$(2.50,2.20,4.44)\%$ of the tree at $n=10$,
$(1.34,0.918,2.65)\%$ at $n=12$, and
$(0.529,0.401,1.53)\%$ at $n=14$, respectively.
Paired 95\% intervals favor IQO over random and DEC over IQO
at every size on this sparse ensemble.

\subsection{Classical performance context}
\label{app:classicalbaselines}
Table~\ref{tab:hybridcontrol} holds the spectral bound fixed and
compares finite-shot Tree-IQO with a classical DEC schedule and
leaf, and with DEC search supplemented by the already generated
eight-shot $p=1$ IQO candidates.
DEC visits fewer nodes than this $p=1$ Tree-IQO realization on
$66\%,88\%,94\%$ of matched instances at $n=10,12,14$.
The IQO sample incumbent beats the DEC leaf on $34\%,32\%,28\%$
of them; adding the candidates to DEC gives the modest mean
improvements in Table~\ref{tab:hybridcontrol}, with paired 95\%
intervals below zero at each size. The hybrid uses extra quantum
work, so this is not a runtime advantage. Stronger classical
MaxCut heuristics and exact solvers exist~\cite{RendlRinaldiWiegele2010,KrislockMalickRoupin2017,Dunning2018}.
We have not made a runtime-matched solver comparison.
\begin{table}[h]
\caption{Mean fraction of the full tree visited before exact
certification on matched complete weighted MaxCut instances.
DEC + IQO samples keeps DEC ordering and the spectral bound;
only its incumbent pool receives the additional IQO candidates.}
\label{tab:hybridcontrol}
\centering
\begin{tabular}{c@{\quad}ccc}
\toprule
$n$ & Tree-IQO+spec. & Classical DEC & DEC+IQO samples \\
\midrule
10 & $9.6\%$ & $7.8\%$ & $7.2\%$ \\
12 & $6.8\%$ & $5.1\%$ & $4.9\%$ \\
14 & $5.0\%$ & $3.3\%$ & $3.2\%$ \\
\bottomrule
\end{tabular}
\end{table}
\end{document}